\documentclass[a4paper,fleqn,usenatbib]{mnras}
\usepackage{graphicx}
\usepackage[english]{babel}
\usepackage{float}
\usepackage{caption}
\usepackage{subcaption}
\usepackage[authoryear]{natbib}
\usepackage{color}
\usepackage{amsmath}
\usepackage{amssymb}
\usepackage{cleveref}
\usepackage[utf8]{inputenc}
\crefname{section}{§}{§§}
\Crefname{section}{§}{§§}

\title[Photo-ionization of planetary winds]{Photo-ionization of planetary winds: case study HD 209458b}
\author[Schneiter et al.]{E. M. Schneiter,\thanks{E-mail:matias@oac.uncor.edu}$^{1,2,3}$
  A. Esquivel,$^{4}$ C. S. Villarreal D'Angelo,$^{1}$ \newauthor
 P. F. Vel\'azquez,$^{4}$ A. C. Raga$^{4}$ and A. Costa$^{1}$ \\
$^{1}$Instituto de Astronom\'ia Te\'orica y Experimental, Universidad Nacional de C\'ordoba, C\'ordoba, Argentina\\
$^{2}$Departamento de Materiales y Tecnolog\'ia, UNC, C\'ordoba, Argentina\\
$^{3}$Departament of Astronomy, AlbaNova, Stockholm University, Sweden \\
$^{4}$Instituto de Ciencias Nucleares, Universidad Nacional Aut\'onoma
de M\'exico, M\'exico D. F. M\'exico\\} 

\date{Accepted. Received; in original form}

\pubyear{2015}

\begin{document}
\pagerange{\pageref{firstpage}--\pageref{lastpage}}

\maketitle

\label{firstpage}

\begin{abstract}
  Close-in hot Jupiters are exposed to a tremendous photon flux that
  ionizes the neutral escaping material from the planet leaving an
  observable imprint that makes them an interesting laboratory for
  testing theoretical models. In this work we present 3D hydrodynamic
  simulations with radiation transfer calculations of a close-in
  exoplanet in a blow-off state. We calculate the Ly-$\alpha$
  absorption and compare it with observations of HD 209458b and
  previous simplified model results.
  Our results show that the
  hydrodynamic interaction together with a proper calculation of the
  photoionization proccess are able to reproduce the main features of
  the observed Ly-$\alpha$ absorption, in particular at the
  blue-shifted wings of the line.
  We found that the ionizing stellar flux produce an almost linear
  effect on the amount of absorption in the wake. Varying the
  planetary mass loss rate and the radiation flux, we 
  were able to reproduce the $10\%$ absorption observed at
  $-100~\mathrm{km~s^{-1}}$.

\end{abstract}


\section{Introduction}

The discovery of the first hot Jupiter transiting its host star HD
209458 took place more than a decade ago
\citep[see][]{charbonneau2000, henry2000}. The analysis of data
samples of the extra-solar planet transit HD 209458b obtained with the
Space Telescope Imaging Spectrograph on board of the Hubble Space
Telescope made by \citet{vidal2003} revealed the existence of an
extended upper atmosphere, and opened a discussion on whether the
atmosphere is being lost \citep[see][]{benjaffel2007, benjaffel2008}.
\citet{vidal2008} revisited the observations and the theoretical- works
based on the picture of an escaping atmosphere
\citep[e.g.][]{lammer2003, lecavelier2004, yelle2004, yelle2006,
  baraffe2005, tian2005, schneiter2007, garcia2007} and confirmed it.
Since then numerous theoretical works have been carried out shifting
the discussion to the escaping mechanism, putting the focus on trying
to explain the absorption produced at high velocities
\citep[see][]{holmstrom2008, ekenback2010, tremblin2013,
  villarreal2014, bourrier2013}. \citet{vidal2008} evaluated
the absorption seen during transit within the `blue' and the `red'
sides \citep[of the wavelength domain defined in][]{benjaffel2007}
finding a $9.8\pm 1.8\%$ and $5.2\pm 1.0\%$ absorption, respectively.

\citet{vidal2003} first interpreted the absorption as hydrogen atoms
in the exosphere undergoing hydrodynamic escape, accelerated by the
stellar radiation pressure.  Later, \citet{schneiter2007} developed a
3D hydrodynamic model of the interaction between the stellar wind and
the gas escaping from the planet (HD 209458b). By comparing the
calculated Ly-$\alpha$ absorption with the observations they were
able to estimate an upper limit for the planetary mass loss rate
($\dot{M}_\mathrm{p}$).  In \citet{villarreal2014} the stellar wind conditions
were improved by comprehensively studying the stellar wind
parameters. This was done by using a polytropic model to estimate the
initial conditions for the stellar wind at the radial distance (from
the centre of the star) at which it was imposed. Several stellar wind
velocities where tested together with the coupled stellar wind
temperature ($T_*$), finding a suitable range for $\dot{M}_\mathrm{p}$
($[3$--$5] \times 10^{10}$ g s$^{-1}$).
\citet{schneiter2007} and
\citet{villarreal2014}, distinguished the planetary wind from the
stellar wind with the aid of a passive scalar, and calculated the
Ly-$\alpha$ absorption assuming that all neutral material with
temperatures above $10^5~$K was ionized. These works were able to
reproduce the velocity range of neutral planetary atoms responsible
for the observed absorption in the Ly-$\alpha$ line.

The work done in \citet{holmstrom2008} and its follow up
\citet{ekenback2010} reproduce the absorption observed at Doppler
shift $-100~$km~s$^{-1}$. They simulated the production of energetic
neutral hydrogen (ENAs) through charge exchange between stellar wind
protons and exospheric hydrogen, assuming stellar wind conditions
similar to those at an equivalent distance in our Sun.
\citet{tremblin2013} further explored the charge-exchange scenario by
means of a 2D hydrodynamic model, making emphasis on the mixing
layer between both winds.  

\citet{bourrier2013} presented a 3D Monte-Carlo numerical particle
model developed to simu-late the escaping gas of an exoplanet. In their
model the focus was set on the radiation pressure, taking into account
the ionization, the self-shielding, and the stellar wind
interaction. The charge-exchange process was included as an impulse
given by the incoming stellar wind proton, aiming, this way, to explain
the blue-shifted absorption. 

Even though most of these models employ different acceleration
mechanism for the neutral particles, they all have been capable of
reproducing the global structure of the upper atmosphere of exoplanets
undergoing mass loss due to Roche lobe overflow, predicting planetary
mass loss rates in the range $10^9$--$10^{11}$ g s$^{-1}$, leaving the
question of which particle acceleration mechanism is responsible for
the absorption at high Doppler shift, still open.

Recent progress has been made to model the planetary
  mass loss rate including the effect of magnetic fields in both the
  star and the planet. It has been shown for instance that space
  weather events (such as CME) or the parameters of the planetary
  system itself, can have an influence on the mass loss from the planet \citep[see for
  example][]{cohen2011a,cohen2011b, owen2014, trammell2014, adams2011,
    matsakos2015}. \citet{owen2014} find, through a number of
  numerical simulations that include the stellar UV flux, that the
  presence of magnetic fields, can reduce the amount of planetary mass
  loss rate by aproximately an order of magnitude if the planet can
  sustain moderate magnetic fields values ($B_p>1~G$). Finally, as
  mentioned in \citet{cohen2011a,cohen2011b} and in \citet{owen2014},
  the interaction of the planetary magnetosphere with the magnetized
  stellar wind is expected to be highly dynamic and change with time
  according to the parameters of the planetary system. 

Here we present a 3D numerical model of the hydrodynamic interaction
between the winds that includes in a self-consistent manner the
calculation of the photo-ionization, and we focus in the effects of
varying the ionizing flux, the stellar wind parameters and the mass
loss rate of the planet.
Similar simulations have recently been presented by
\citet{tripathi2015}, where the evaporation of the exoplanet
atmosphere is modeled with radiation hydrodynamics. In such work, an
asymmetric planetary wind is produced by a plane parallel radiative
field and stellar wind. However, their focus is in the mass loss from
the planet, and not on the extended wake.

The model, some details of the code, and the parameters employed are
presented in Section \ref{introduction}.  
The results are presented in Section \ref{results} and
discussed in Section \ref{discussion}. The concluding remarks are
given in Section \ref{conclusions}.

\section{The model}\label{introduction}

We used the radiation-hydrodynamics code {\sc guacho}
\citep{esquivel2009,2013ApJ...779..111E} to produce synthetic
Ly-$\alpha$ absorption maps of a hot-Jupiter around a solar type star
(i.e. the HD 209458 system).

\subsection{The hydrodynamics core}\label{sec:code}

The code solves the ideal hydrodynamics equations, along with
gravity and radiative gains and losses, in a Cartesian grid:
\begin{equation}
\frac{\partial \rho}{\partial t} + \nabla \cdot (\rho {\bf u})=0,
\label{eq:cont}
\end{equation}

\begin{equation}
\frac{\partial (\rho {\bf u})}{\partial t} + \nabla \cdot (\rho {\bf
  uu}+{\bf I}P)=\rho {\bf g},
\label{eg:mom}
\end{equation}

\begin{equation}
\frac{\partial E}{\partial t} + \nabla \cdot [{\bf
    u}(E+P)]=G_\mathrm{rad}-L_\mathrm{rad}+ \rho \left( {\bf g} \cdot
     {\bf u}\right),
\label{eq:ener}
\end{equation}
where $\rho$, {\bf u}, P, and E are the mass density, velocity,
thermal pressure and energy density, respectively. {\bf I} 
is the identity matrix, $\mathbf{g}$ is the acceleration due to
gravitational forces, while $G_\mathrm{rad}$ and $L_\mathrm{rad}$ the
gains and losses due to radiation.
The total energy density and thermal pressure are related by an ideal
gas equation of state $E = \rho \vert{\bf u}\vert^2/2 + P/(\gamma-1)$,
where $\gamma=5/3$ is the ratio between specific heat capacities.

The hydrodynamics equations (left hand side of Equations
\ref{eq:cont}--\ref{eq:ener}), are advanced with a second order
Godunov method with an approximate Riemann solver (HLLC,
\citealt{torobook}), and a linear reconstruction of the primitive
variables using the $\mathrm{minmod}$ slope limiter to ensure stability.

\subsection{Source terms}

At every time-step, the hydrodynamic variables are updated, with the
updates values the source terms (right hand side of Equations
\ref{eq:cont}--\ref{eq:ener}) are computed and added to the solution
in a semi-implicit scheme.

\subsubsection{Gravity}
For the gravitational force we add to each cell on the grid the
acceleration due to two point masses, the mass of the planet (whose
position changes as it orbits around the star), and the mass of star.
The mass of the star is reduced to $1/3$ of the actual value for the
computation of the gravitational acceleration, to emulate the effect
of the radiation pressure force that acts in opposition to gravity
\citep[see for example][]{vidal2003}.  

A more detailed calculation of the radiation pressure is possible.
For instance, \citet[][]{bourrier2013, bourrier2015}, calculated the
ratio of the radiation pressure to gravitational forces (the $\beta$
parameter in their models, which is proportional to the Ly-$\alpha$
flux) to model the evaporation of the atmospheres of HD 209458b and HD
189733b. Such calculation involves the knowledge of the Ly-$\alpha$
emission as seen by each cell, which is beyond the scope of the
present paper. We plan to pursue this refinement in a future study.

\subsubsection{Heating, cooling, and radiative transfer}

In order to consider the photoionization of H, together with the
gas-dynamic equations we integrate an additional equation for neutral
hydrogen of the form:
\begin{equation}
 \begin{split}
  \frac{\partial n_\mathrm{HI}}{\partial t} + \nabla.(n_\mathrm{HI}
  {\bf{u}})=  & (n-n_\mathrm{HI})^2\alpha(T)\\
& -(n-n_\mathrm{HI})n_\mathrm{HI}c(T)-n_\mathrm{HI}\phi,
 \end{split}
\label{eq:hrate}
\end{equation}
where $\alpha (T)$ is the
recombination coefficient, $c(T)$ is the collisional ionization
coefficient of hydrogen, and $\phi$ the photoionization rate.
In Equation (\ref{eq:hrate}) we have assummed that the electron
density is equal to that of ionized hydrogen
${n_e=n_\mathrm{HII}=n-n_\mathrm{HI}}$, where $n_\mathrm{HI}$ and $n$ are the
neutral and total hydrogen density, respectively.

The hydrogen photoionization rate ($\phi$), due to the UV photons emitted
by the star is given by: 
\begin{equation}
\phi = \int^{\infty}_{\nu_0} \frac{4\pi J_{\nu_0}}{h \nu} a_{\nu} d\nu
\end{equation}
where $J_{\nu_0}$ is the solid angle averaged intensity of the
ionizing radiative field, $\nu_0$ is the Lyman limit frequency, $h$ is
Plank's constant and $a_\nu$ is the photoionization cross section of H.
To compute the photoionization rate we have used the standard `grey
approximation' in which the frequency dependence of the cross section
is not considered, and all the photons are assumed to be at the Lyman
limit, with a constant photoionization cross section $a_0=6.3\times
10^{-18}~\mathrm{cm^{-2}}$.

To include the photo-ionizing radiation of the host star we modified
the ray tracing method described in \citet{2013ApJ...779..111E}. In
the version used here we divide the $EUV$ stellar luminosity into
$10^7$ photon packets, which are launched in random directions, from
random positions at the surface of the star.
\footnote{We have made a convergence test to ensure that this number
  of photon packets is sufficient. We repeated the same calculation
  for a model (B2b) with $10^5$, $10^6$, $10^7$ and $5\times 10^7$
  photon packets. We then computed
the Ly-$\alpha$ absorption profile. With $10^7$ rays the results
were almost indistinguishable from those with a larger number.}
During the propagation,
each photon-package is decimated by factors of $e^{-\Delta \tau}$ when
travelling through the intercepted cells (when neutral material is
encountered, $\Delta \tau = a_0 n_\mathrm{HI} \Delta l$, where $\Delta
l$ is the path-length traversed).
 The photoionizing rate is obtained by
equating the photon rate to the ionizations per unit time in each
cell, as the photon packets reach it
\begin{equation}
\label{eq:rate-phi}
S=n_\mathrm{H_I}\,\phi\,dV.
\end{equation}

The contributions from the absorbed
photons are added to the photoionization (Eq. \ref{eq:hrate}), and
heating rates of each cell in the computational grid.  The resulting
heating rate
\begin{equation}
  \label{eq:heat}
  \psi=n_\mathrm{HI}\phi E_0, 
\end{equation}
is then included in the energy term (Eq. \ref{eq:ener}) after each
hydrodynamical timestep. The energy gain per ionization is assumed to
be $E_0=13~\mathrm{eV}$. 
The cooling rate at low temperatures ($< 3 \times 10^4~\mathrm{K}$), is
due to collisional ionization of oxygen, assuming that the ionization
of \ion{O}{II} follows that of \ion{H}{II} (obtained from
Eq. \ref{eq:hrate}). This is a good approximation due to the
efficient charge exchange between H and O, see \citealt{1993ApJ...409..705H}.
At higher temperatures, when oxygen is expected to be more than singly
ionized, the cooling is switched to a coronal equilibrium cooling curve.

\subsection{Parameters of the simulations}

The star/planet system is modeled as the interaction of two isotropic
wind sources in orbit.  There are many physical
  processes that could lead to an anisotropic planetary wind. For
  instance, a tidally locked planet would lead to the same hemisphere
  facing the host star, and, unless there are atmospheric flows that
  transport heat efficiently \citep[see][for a discussion on internal
    flows]{batygin2014, showman2008} that are capable of thermalising
  the whole atmosphere, the winds will be assymetric. This effect has
  been studied before in \citet{villarreal2014} by means of a few toy
  models. In that work, inspite of the simplicity of the models, only
  a minor effect in the wake structure that produced most of the
  Ly-$\alpha$ absorption was found. Recently, a more detailed
  calculation of the photoevaporation of the planet atmosphere was
  presented in \citet{tripathi2015}. They used 3D
  radiation-hydrodynamics and arrive to the same value of
  $\dot{M}_\mathrm{p}$ that we adopt ($2\times
  10^{10}\mathrm{g~s^{-1}}$). For simplicity, and left to a future
  work, we assume isotropic planetary and stellar winds.

We place the  source that corresponds to the star at
the center of the computational domain, which coincides with the
origin of a Cartesian grid. The planet orbits the star in the
$xz$-plane in an anti-clockwise direction.
Both winds are reimposed at every time-step with the planet position
updated according to its orbital period
($\tau_\mathrm{orb}=3.52~\mathrm{days}$). 
The orbit is assumed circular with a radius of $0.047~\mathrm{AU}$.
The initial position of the planet is $25\degr$ `behind' the
$x$-axis to ensure that the wake is formed by 
the time it reaches the $z$-axis, wich we have taken to be 
the line of sight (LOS), unless otherwise stated.

We ran a total of 19 models, varying  the stellar wind temperature and
velocity, the photoionizing rate and the mass loss rate of the planet. 
For clarity we present the parameters used in the simulations in two
tables. In Table \ref{tab:1} we show the star and planet parameters,
and the symbols we use for them. In this table we show the ranges of
the parameters that we vary as well as some common quantities in the
simulations. In Table \ref{tab:2} we show the details of each of the
models.

\begin{table}
\caption{Stellar and planetary winds parameters employed in simulations}
  \label{tab:1}
  \begin{tabular}{lll}
\hline
   {\bf Stellar parameters} & Symbol & Value 
    \\ \hline
   Radius               & $R_*$  &   $1.146~\mathrm{R_\odot}$    \\
   Mass                 & $M_*$  &   $1.148~\mathrm{M_\odot}$    \\
   Wind velocity        & $v_\mathrm{*}$  &   $130$--$372~\mathrm{km~s^{-1}}$\\
   Wind launch radius   & $R_\mathrm{w,*}$  &   $3.5$--$6.9~R_*$    \\
   Wind temperature     & $T_\mathrm{*}$  &   $1$--$3~\mathrm{MK}$               \\
   Mass loss rate       & $\dot{M_*}$ &
                                                 $2.0~\mathrm{M_{\odot}~yr^{-1}}$
    \\
   Ionizing photon rate & $S_0$ & $2.4\times 10^{38}~\mathrm{s^{-1}}$  \\
   \hline
   {\bf Planetary  parameters} & Symbol  & Value  
    \\ \hline
   Radius               & $R_\mathrm{p}$ &   $1.38~\mathrm{R_{Jup}}$    \\
   Mass                 & $M_\mathrm{p}$ &   $0.67~\mathrm{M_{Jup}}$    \\
   Wind velocity        & $v_\mathrm{p}$ &   $10~\mathrm{km~s^{-1}}$  \\
   Wind launch radius   & $R_\mathrm{w,p}$ &   $3~R_\mathrm{p}$              \\
   Wind temperature     & $T_\mathrm{p}$ &   $1 \times 10^4~\mathrm{K}$    \\
   Mass loss rate       & $\dot{M}_\mathrm{p}$ & $(1$--$2)
                                                             \times
                                                             10^{10}~\mathrm{g~
                                                             s ^{-1}}$ \\
    \hline
  \end{tabular}
  
\end{table}

\begin{table*}
\begin{minipage}{100mm}
\caption{Parameters of all the simulations.}
  \label{tab:2}
  \begin{tabular}{lccccc}
\hline
    Runs & $\dot{M}_\mathrm{p}$  & 
$v_*$ &  
Grid\footnote{All simulations have the same resolution, but to avoid
              neutral material from the tail leaving the
              domain we had to increase the grid size in some of the runs.}
& $T$ & Photon Rate  \\
 & $(\times 10^{10}~\mathrm{g~s^{-1}})$ & $(\mathrm{km~s^{-1}})$& $(\mathrm{AU})$ &
 $(\times 10^6~\mathrm{K})$ & $(\mathrm{s^{-1}})$
    \\ \hline
A1a  &  $1$  & $130$  & $0.2\times0.05\times0.2$ & $1$    & $0.2\times S_0$ \\
A1b  &  $2$  & $130$  & $0.2\times0.05\times0.2$ & $1$    & $0.2\times S_0$ \\
A2a  &  $1$  & $130$  & $0.2\times0.05\times0.2$ & $1$    & $S_0~~~~~~~~$ \\    
A2b  &  $2$  & $130$  & $0.2\times0.05\times0.2$ & $1$    & $S_0~~~~~~~~$ \\
ANRb &  $2$  & $130$  & $0.2\times0.05\times0.2$ & $1$    &  No Rad~~\\
A3a  &  $1$  & $130$  & $0.2\times0.05\times0.2$ & $1$    & $5\times S_0~~~$ \\
A3b  &  $2$  & $130$  & $0.2\times0.05\times0.2$ & $1$    & $5\times S_0~~~$ \\
B1a  &  $1$  & $205$  & $0.2\times0.05\times0.2$ & $1.3$  & $0.2\times S_0$ \\
B1b  &  $2$  & $205$  & $0.2\times0.05\times0.2$ & $1.3$  & $0.2\times S_0$ \\
B2a  &  $1$  & $205$  & $0.2\times0.05\times0.2$ & $1.3$  & $S_0~~~~~~~~$ \\
B2b  &  $2$  & $205$  & $0.2\times0.05\times0.2$ & $1.3$  & $S_0~~~~~~~~$ \\
B3a  &  $1$  & $205$  & $0.2\times0.05\times0.2$ & $1.3$  & $5\times S_0~~~$ \\
B3b  &  $2$  & $205$  & $0.2\times0.05\times0.2$ & $1.3$  & $5\times S_0~~~$ \\
C1a  &  $1$  & $372$  & $0.3\times0.075\times0.3$ & $3$    & $0.2\times S_0$ \\
C1b  &  $2$  & $372$  & $0.3\times0.075\times0.3$ & $3$    & $0.2\times S_0$ \\
C2a  &  $1$  & $372$  & $0.3\times0.075\times0.3$ & $3$    & $S_0~~~~~~~~$ \\
C2b  &  $2$  & $372$  & $0.3\times0.075\times0.3$ & $3$    & $S_0~~~~~~~~$ \\
C3a  &  $1$  & $372$  & $0.3\times0.075\times0.3$ & $3$    & $5\times S_0~~~$ \\
C3b  &  $2$  & $372$  & $0.3\times0.075\times0.3$ & $3$    & $5\times S_0~~~$ 
    \\ \hline
  \end{tabular}
\end{minipage}
\end{table*}

The stellar wind is imposed fully ionized near the sonic point with
the corresponding temperature and velocity, the density follows an
$\propto r^{-2}$ profile, scaled to give the corresponding mass loss
rate. The temperature and velocity of the winds were extrapolated
from the surface of the star to the launch position with the aid of a
polytropic model with index close $1.01$
\citep[see][]{villarreal2014}.  

The planetary wind is imposed with an ionization
  fraction of $0.8$, at $3~R_\mathrm{p}$ (roughly the Hill
  radius). This choice is based on the results obtained in
  \citet{murray-clay2009}, where a one-dimensional model for
  photoevaporative mass loss from a hot Jupiter is constructed. They
  focus on the escape of hydrogen originated on the substellar point of
  the planet, and assume that mass loss occurs in the form of a
  steady, hydrodynamic, transonic wind.  Their results predict that
  $20\%$ of the material remains neutral at $3~R_\mathrm{p}$. This
  value, however, depends strongly on the model
  assumptions. For instance, \citet{koskinen2013} predicts a
  transition \ion{H}{i}/\ion{H}{ii} at an altitude of $3.1$R$_p$, implying that $50\%$
  of the escaping material remains neutral at this
  altitude. The mass loss rate adopted is also consistent with that
  obtained in \citet{tripathi2015}, who presented 3D simulations of
  the photoevaporation near the planet atmosphere.

In our models, the temperature and terminal velocity
are set constant at the base of the planetary wind, while the
density profile is scaled to give the desired mass loss rate.

Based on luminosity estimated for HD 209458 in
\citet{sanz-forcada2011} (they find $\log L_\mathrm{EUV} <
27.74~\mathrm{erg~s^{-1}}$) we have chosen a $S_{0}=2.46\times
10^{38}~\mathrm{s^{-1}}$ for the ste-llar photon rate, which
corresponds to a flux of $F_0=884~\mathrm{erg~cm^{-2}~s^{-1}}$ at the
orbital distance of HD 209458b.  Other works, where focus is set on
the escape process which utilize the EUV and X-ray as energy input
consider a range in luminosities/fluxes. For instance,
\citet{tian2005} use the current Solar luminosity value, yielding a
flux of $F=0.15~\mathrm{erg~cm^{-2}~s^{-1}}$ at 1 AU (roughly a tenth
of ours); while \citet{owen2012} adopt a larger photon rate
($S_*=10^{40}~\mathrm{s^{-1}}$).
\citet{murray-clay2009} use the solar $EUV$ flux of
  $450~\mathrm{erg~s^{-1}}$ at $0.05~\mathrm{AU}$, which corresponds
  to a $S_*=1.4\times 10^{37}~\mathrm{s^{-1}}$. Similar values have
  been adopted in the works of \citet{owen2014,matsakos2015}.
To evaluate the effect of the
ionizing photon rate on the escaping planetary material and somehow
cover the range of values discussed in the literature we use three
different values for the photon flux in our simulations
$(0.2,~1.~,5.)\times S_0$. 

\begin{figure}
  \includegraphics[width=\columnwidth]{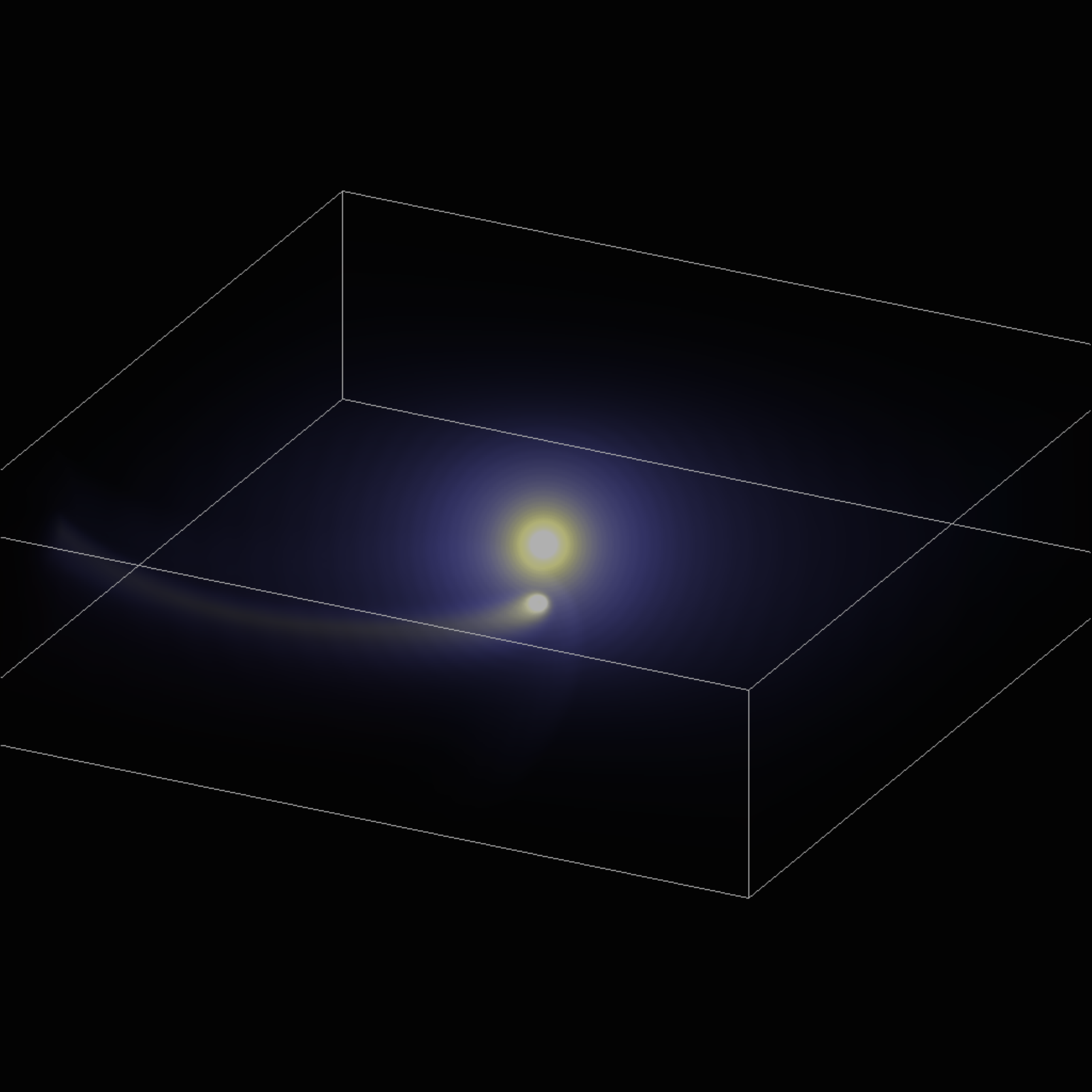}
  \caption{3D rendering of the density of one of the models (C1b, see
    Table \ref{tab:2}), after
    an integration time of $3.8~\mathrm{days}$.}
  \label{fig:render}
\end{figure}

Figure \ref{fig:render} shows the computational domain with a rendering of
the density. The system has reached a fully quasi-stationary state
with a long curved wake behind the planet.
The initial conditions are evolved for an integration time of
$3.8~\mathrm{days}$, in which the planet completes an orbit
around the star.

\section{Results}\label{results}

\begin{figure}
        \centering
        \begin{subfigure}[b]{0.46\textwidth}
                \includegraphics[width=\textwidth]{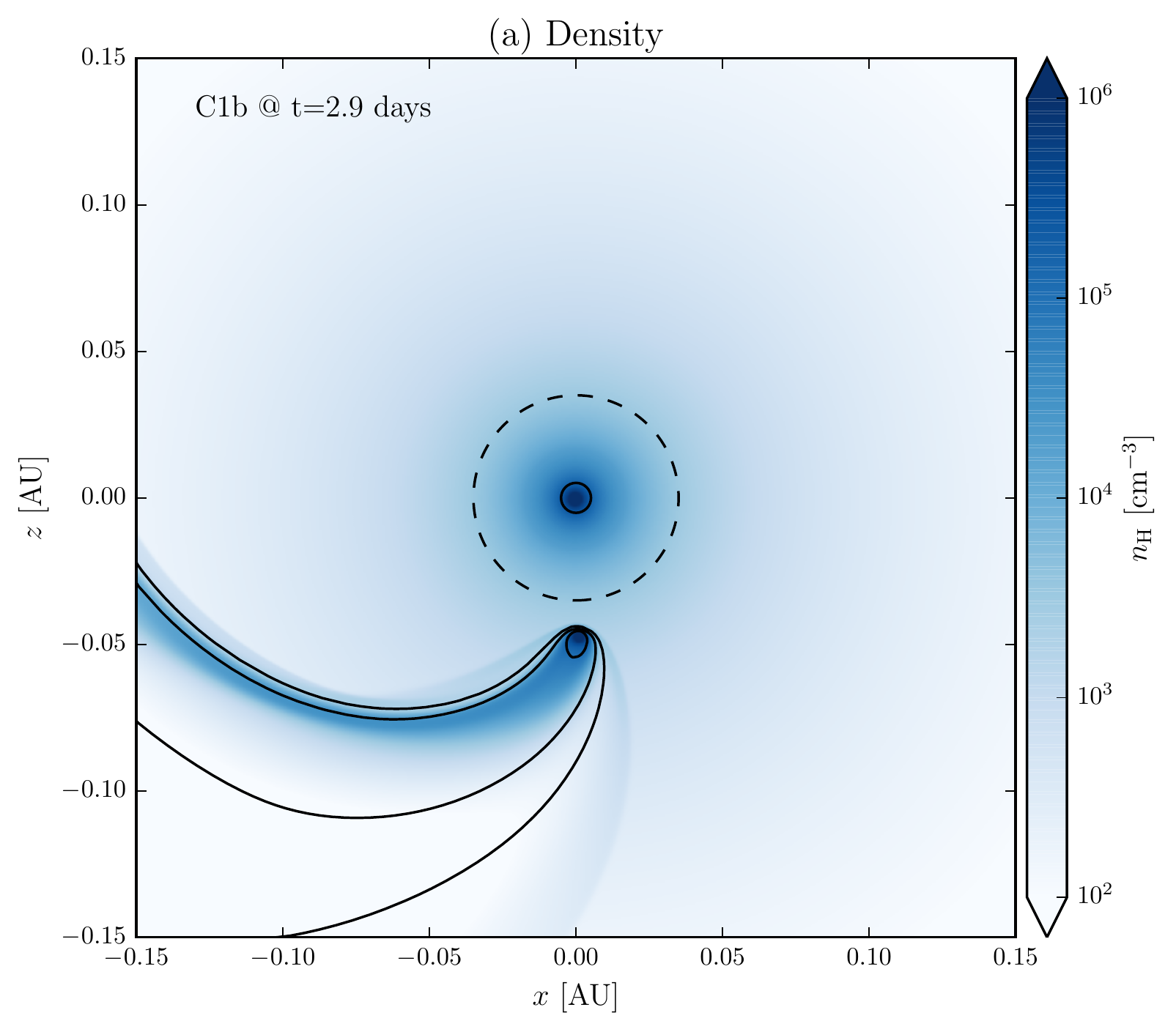}
        \end{subfigure}
        \begin{subfigure}[b]{0.46\textwidth}
                \includegraphics[width=\textwidth]{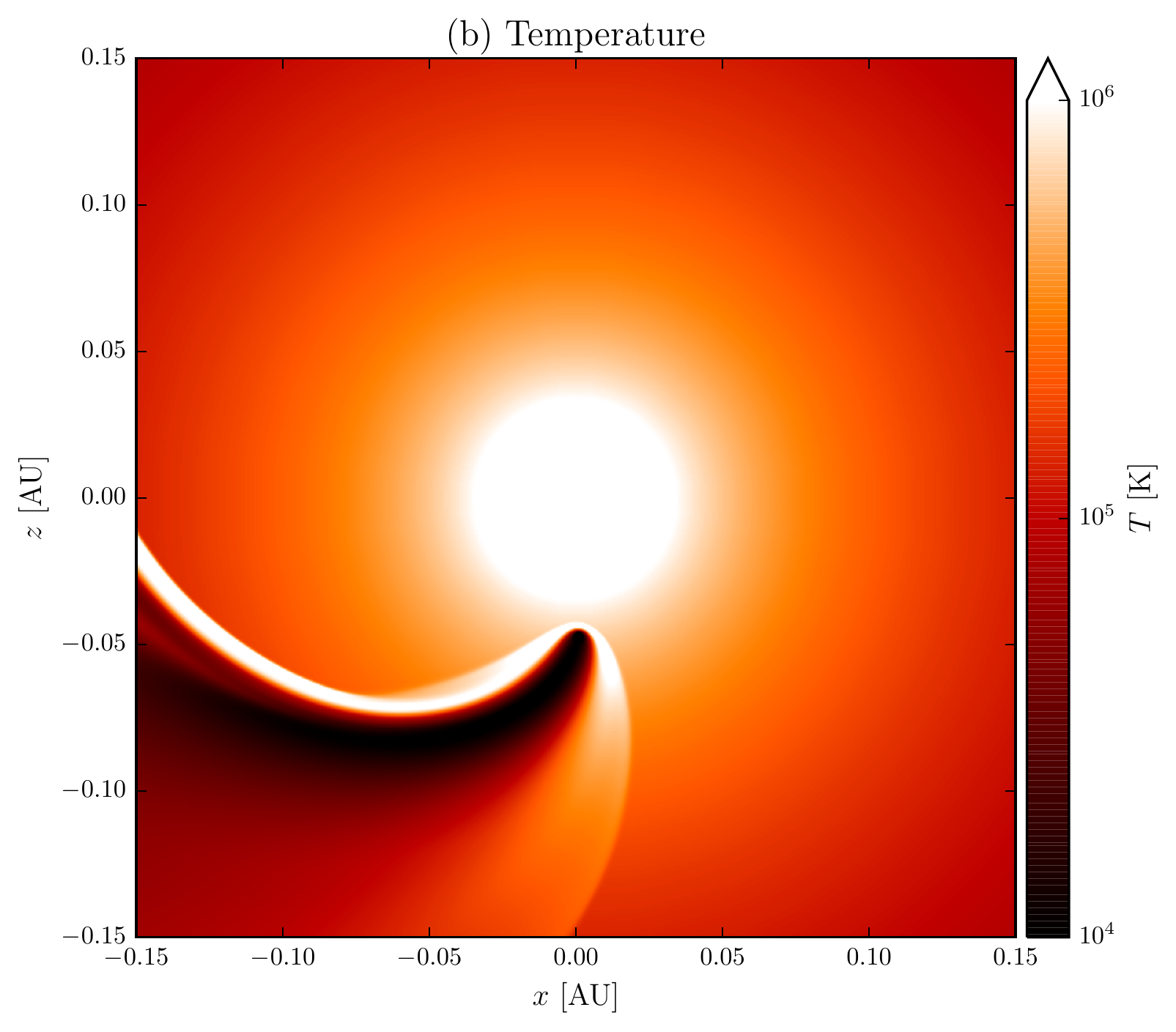}
        \end{subfigure}
        \begin{subfigure}[b]{0.46\textwidth}
                \includegraphics[width=\textwidth]{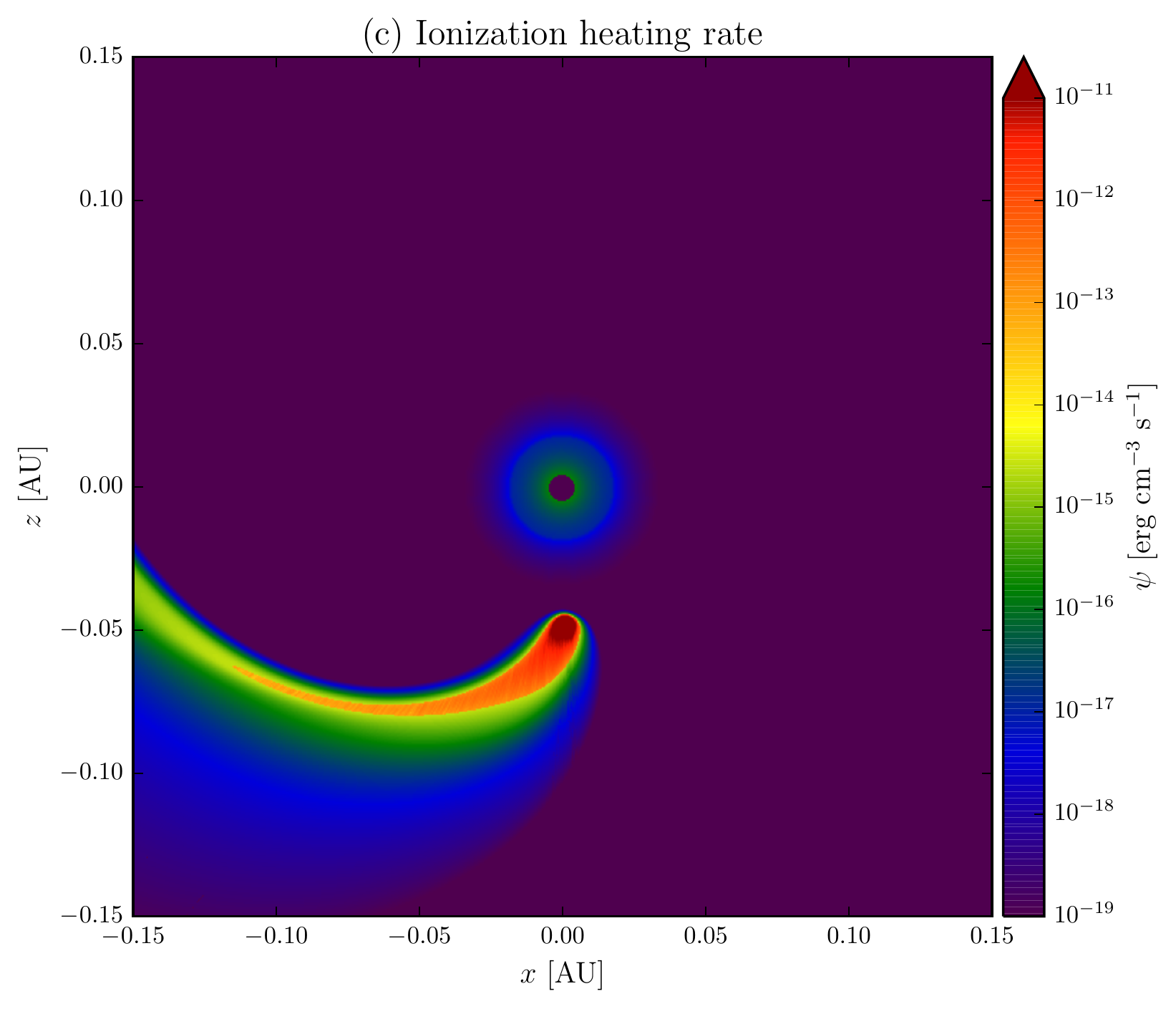}
        \end{subfigure}
        \caption{2D cuts in the orbital plane ($y=0$) of (a) density,
          (b) temperature, and (c) ionization heating rate, at
          $t=2.9~\mathrm{days}$ for model C1b.  In the top panel, we
          show the size of the central star (solid line circle) and
          the size of the region at which the wind is imposed (dashed
          line circle). In the same panel we also show ionization
          fraction iso-contours at values of $0.9$
            (the most interior to the planet), $0.99$, and $0.999$ (the
            most exterior).}
\label{fig:cuts}
\end{figure}

As explored in previous works, the interaction between the stellar and
planetary winds forms a wake structure similar to what is shown in
Figure \ref{fig:render}. The parameters we used for the wind sources
result in similar morphology in all cases, where the morphological
differences lie in the details.  For instance, a rapid wind from the
star results in a less curved and more radial wake than a slow wind,
or a higher mass loss rate of the planet produces a larger bow-shock.
Figure \ref{fig:cuts} shows the density stratification, temperature
and ionizing heating rate for a cut in the orbital plane after an
evolution time of $t=2.9~\mathrm{days}$.

In the top panel we indicate by a solid circle the extent of HD~209458
(the photon packages are launched from this radius), and with a dashed
circle the size of the region in which the stellar wind is imposed.
In addition, the top panel of this Figure includes the ionization
fraction isocontours at $0.9$ (close to the planet),
  $0.99$ and $0.999$ levels. The ionization structure of the tail is
significantly different from our previous models, where the
photoionization was not considered, and a fully neutral planetary wind
was imposed. In \citet{schneiter2007,villarreal2014} we assumed that
material arising from the planetary wind with a temperature
$T<10^5~\mathrm{K}$ was neutral (and it was considered to estimate the
Ly-$\alpha$ absorption). In the models with photoionization we now
assume an ionization fraction of 0.8, which increases rapidly,
implying that there is less material available to absorb.

From the other two panels, temperature and ionization heating rate
stratification  (b and c) it is clear how the impinging photons heat up
the trailing planetary material and how the planet and the high
density interaction region shield the outer part of escaping neutral
atoms in the wake.

\subsection{Ly-$\alpha$ absorption}

To calculate the absorption produced by the escaping hydrogen we
employed a similar post-processing procedure as in
\citet{schneiter2007} and \citet{villarreal2014}. For each of the
simulations we compute the optical depth as a function of velocity
along the LOS
\begin{equation}
  \tau_\mathrm{v_{los}} = \int n_\mathrm{HI} \, a_0 \, \varphi\left( \Delta v_\mathrm{los} \right) ds ,
\label{eq:tau}
\end{equation}
where $\varphi(\Delta v_\mathrm{los})$ is  a Gaussian line-profile.
The optical depth is obtained projecting the velocity on a LOS 
that is almost aligned with  the $z$ and $-z$-axes, inclined
$3.41\degr$ around the $x$-axis to have the same orientation as the 
HD~209548 system seen from Earth (an orbital inclination angle of
$i=86.59\degr$). In the optical depth calculation we have considered a
velocity range of 
$[-300, 300]~\mathrm{km~s^{-1}}$, covered with $250$ velocity bins.
By choosing the LOS close to $z$ and $-z$ we
are able to observe two transits in the same orbit. The optical depth
is only calculated considering the material on the observer's side
from the star. 

\begin{figure}
    \centering
    \includegraphics[width=1\columnwidth]{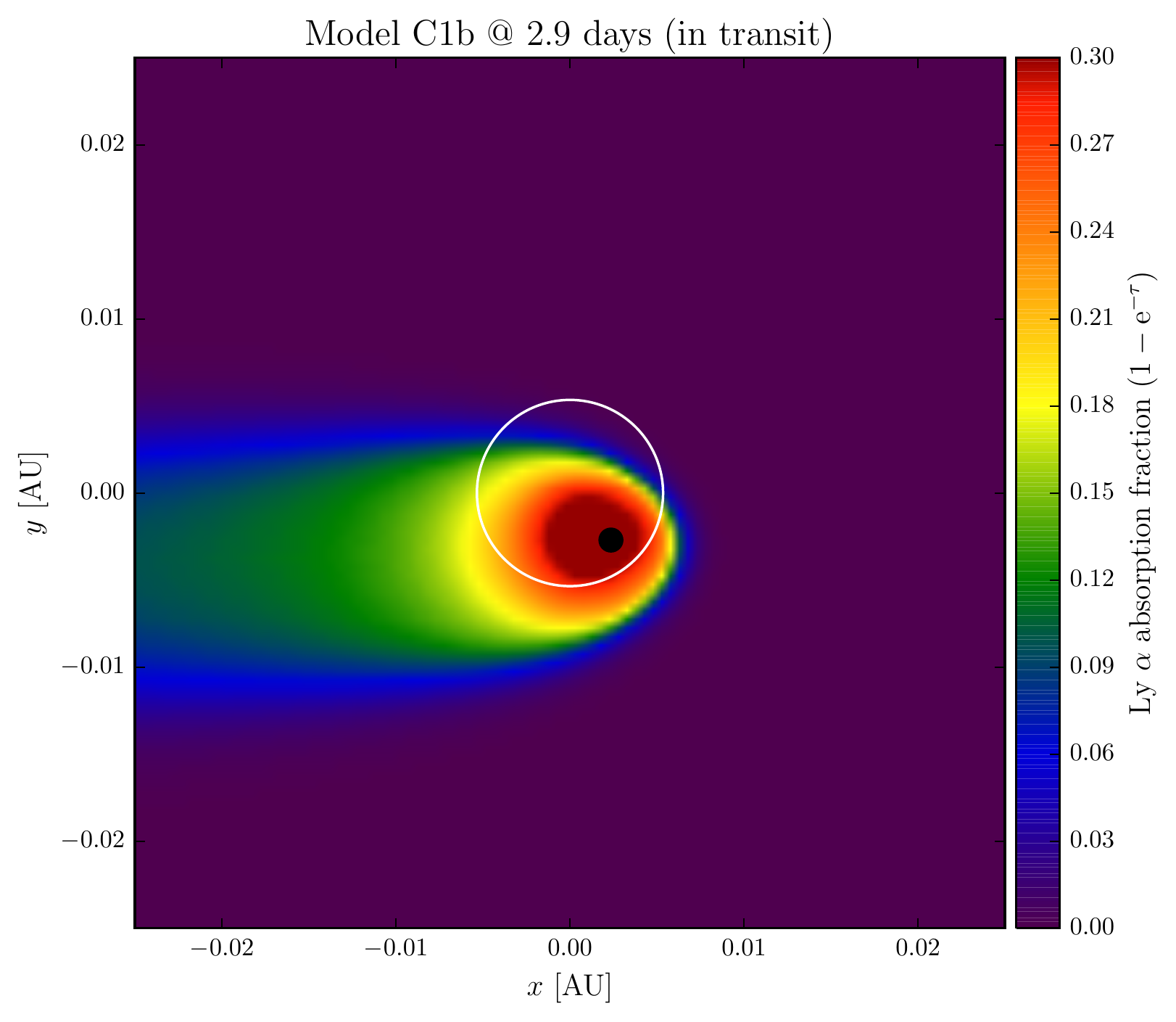}
    \caption{Map of Ly-$\alpha$ optical depth (integrated over all
      velocities) for model C1b in transit
      as observed from the $-z$-axis (integration time of
      $2.9~\mathrm{days}$). The map showed is zoom-in of the entire
      domain. Both the star and the planet are shown in scale by the
      white circle and black disk, respectively. }
    \label{fig:tau}
\end{figure}

The total absorption can be calculated by integrating
$\tau_\mathrm{v_{los}}$ in velocity. The Ly-$\alpha$ emission from the star
would be absorbed by a factor $(1-\mathrm{e}^{-\tau})$, where $\tau$ is
the optical depth integrated over all velocities.
Figure \ref{fig:tau} shows an example  (same model shown in Figures
\ref{fig:render} and \ref{fig:cuts}) of the total  absorption fraction
calculated from the models during a transit, as it would be seen by an
observer in the $-z$ direction (with the inclination of HD 209458b). 

After computing the Ly-$\alpha$ absorption as a function of time, we
found that the wake reaches a quasi-stationary state in only a
fraction of the orbital period. This is somewhat different from our
previous models, where at least half a period was required to reach
such a stationary state. The main reason is that the photoionization
results in a smaller neutral tail.

\begin{figure*}
  \centering
  \begin{subfigure}{0.33\textwidth}
    \includegraphics[width=\textwidth]{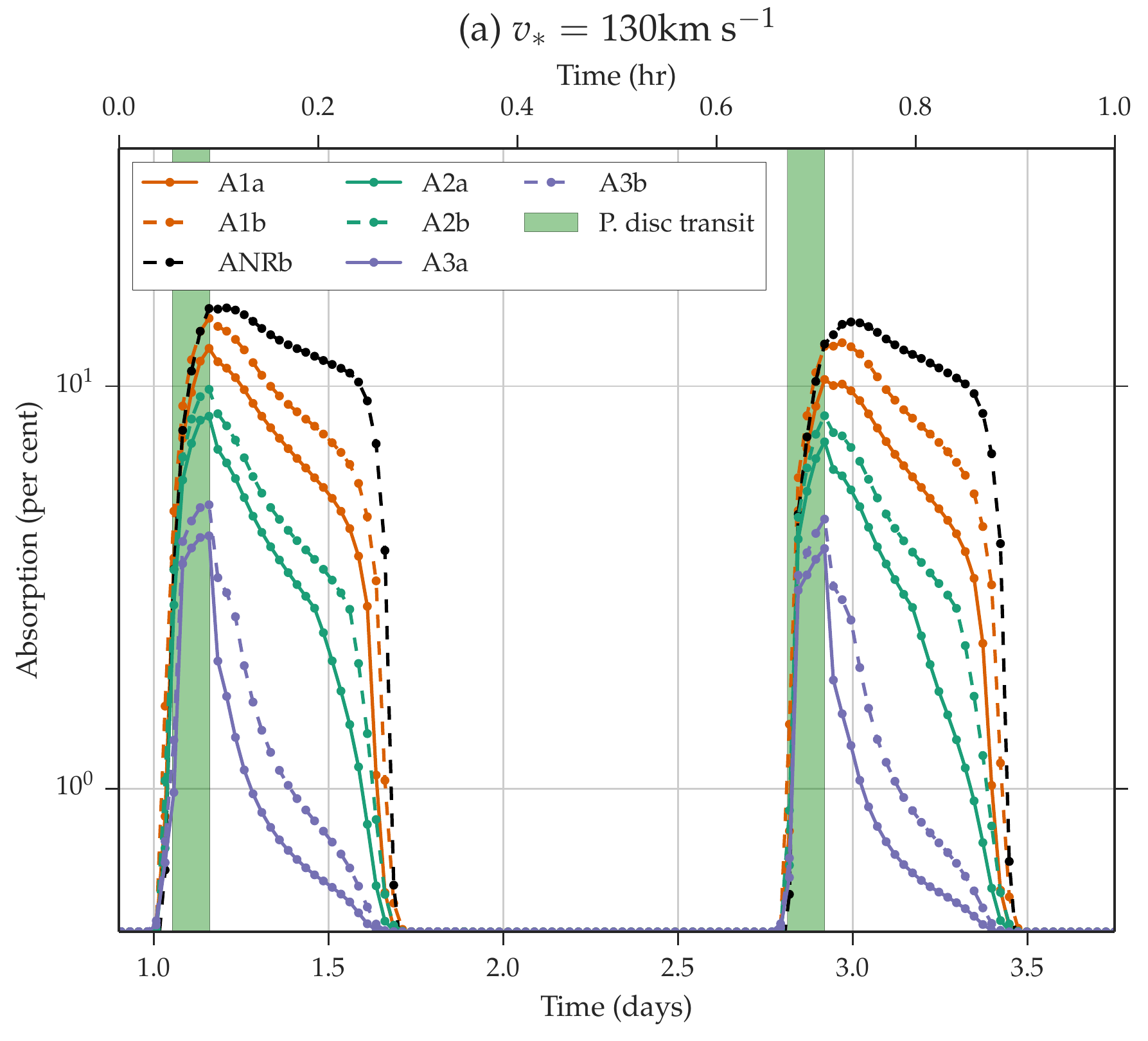}
  \end{subfigure}%
  \begin{subfigure}{0.33\textwidth}
    \includegraphics[width=\textwidth]{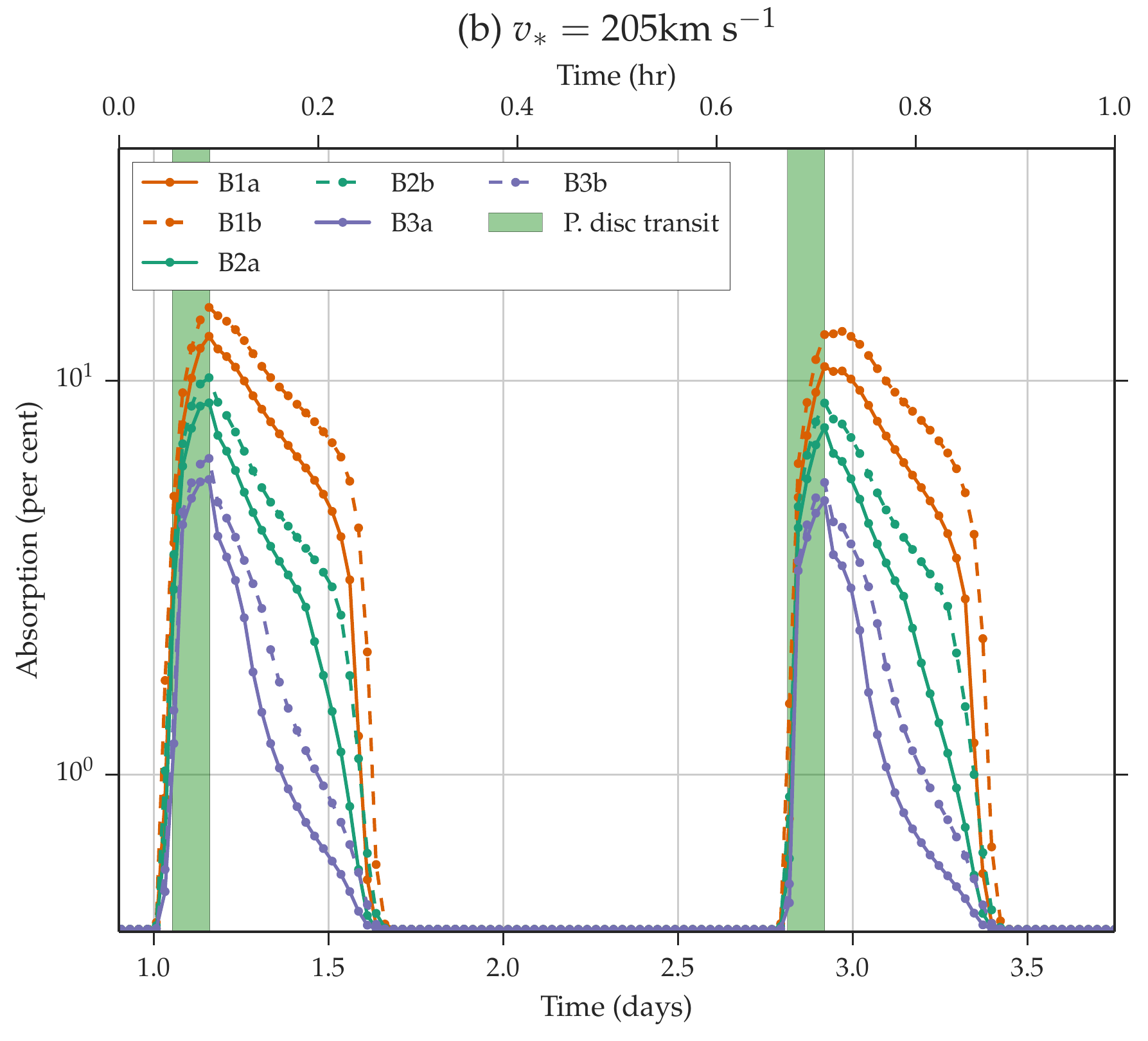}
   \end{subfigure}
  \begin{subfigure}{0.33\textwidth}
    \includegraphics[width=\textwidth]{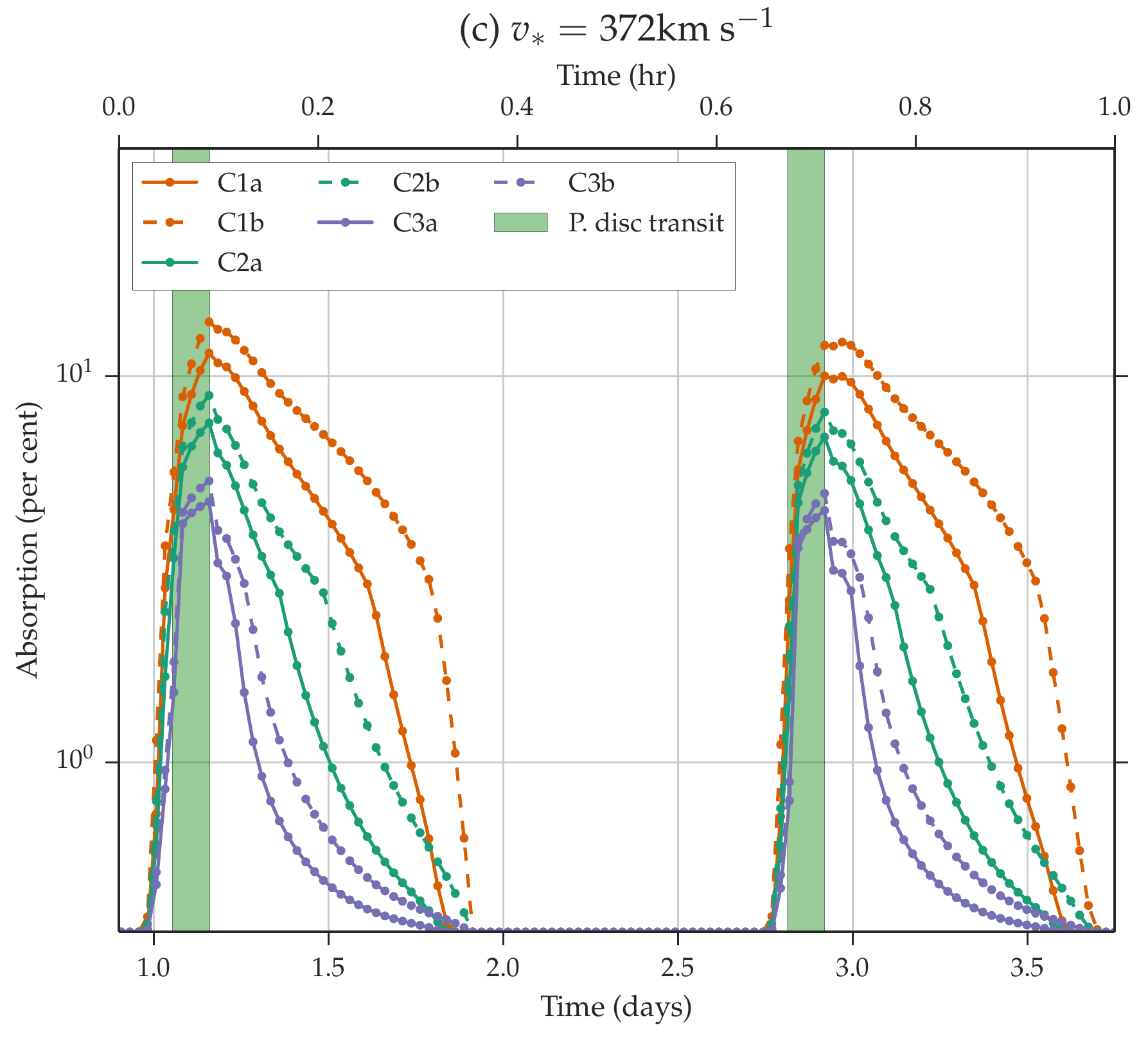}
  \end{subfigure}
  \caption{Velocity averaged total absorption as a function of time
    calculated from an observer situated in the +z (first transit
    $t<2.2~\mathrm{days}$) and -z directions (second transit,
    $t>2.2~\mathrm{days}$). The models are grouped into panels
    according to the stellar wind velocity, as indicated in the title.
    The color coding represents the ionizing flux: $0.2~S_0$ (orange),
    $S_0$ (green), and $5~S_0$ (purple). The dots joined by dotted
    lines correspond to $\dot{M}_\mathrm{p} =2\times
    10^{10}~\mathrm{g~s^{-1}}$, and the solid lines to $\dot{M}_\mathrm{p}
    =1\times 10^{10}~\mathrm{g~s^{-1}}$.
    The green vertical stripe represents the times when a portion of
    the planetary disc blocks the star (i.e. the planet is in transit).}
\label{fig:totabs}
\end{figure*}

Figure \ref{fig:totabs} shows the absorption of the stellar emission
as a function of time, first from the $+z$ direction ($t <
2.2~\mathrm{days}$)  and the subsequent transit as would be measured
from the $-z$ direction (see Figure \ref{fig:cuts}) for all the
models.

We have separated the models according to the stellar wind velocity.
Panel (a) shows all the models with $v_* = 130~\mathrm{km~s^{-1}}$,
including the reference model ANRb that has the same parameters as A2b
(see Table \ref{tab:2} for details) but does not include the
photoionization process.  In this model the absorption was computed as
in \citep{schneiter2007}, that is, assuming that only material
originally from the planet with temperatures below $10^5~\mathrm{K}$
absorb the Ly-$\alpha$ emission.  Panels (b) and (c) show the absorption
for the models with $205$ and $372~\mathrm{km~s^{-1}}$, respectively.
Notice that there are only minor differences in the first and second
transits, showing that the quasi-stationary wake structure forms
rapidly.  The absorption in Figure \ref{fig:totabs} was calculated
excluding the velocity range between $-40$ and
$40~\mathrm{km~s^{-1}}$, which for Earth based observations is
contaminated by the geocoronal emission.
  
From the plots we can see that the maximum absorption changes
when doubling $\dot{M}_{p}$. This is in accordance with the results
obtained in \citet{schneiter2007}and \citet{villarreal2014}.  At the
same time the total absorption is greatly affected by the change in
$F_\mathrm{EUV}$, producing appreciable changes both in the maximum
absorption as well as in the tail of the wake (absorption after the
transit of the planet), a denser and more extended
  structure is formed for low ionizing fluxes.
 
\begin{figure*}
  \centering
  \begin{subfigure}{0.33\textwidth}
    \includegraphics[width=\textwidth]{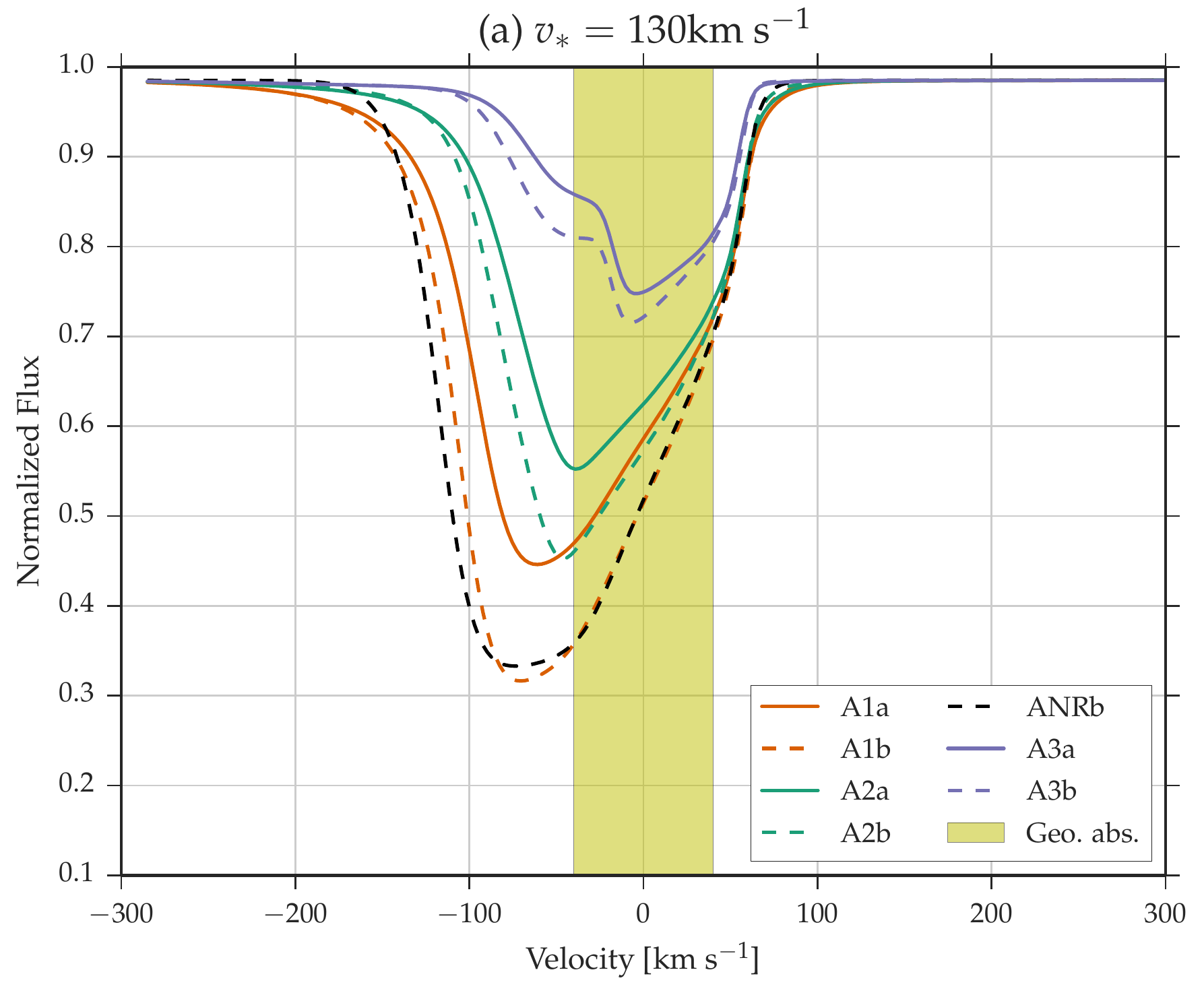}
  \end{subfigure}
  \begin{subfigure}{0.33\textwidth}
    \includegraphics[width=\textwidth]{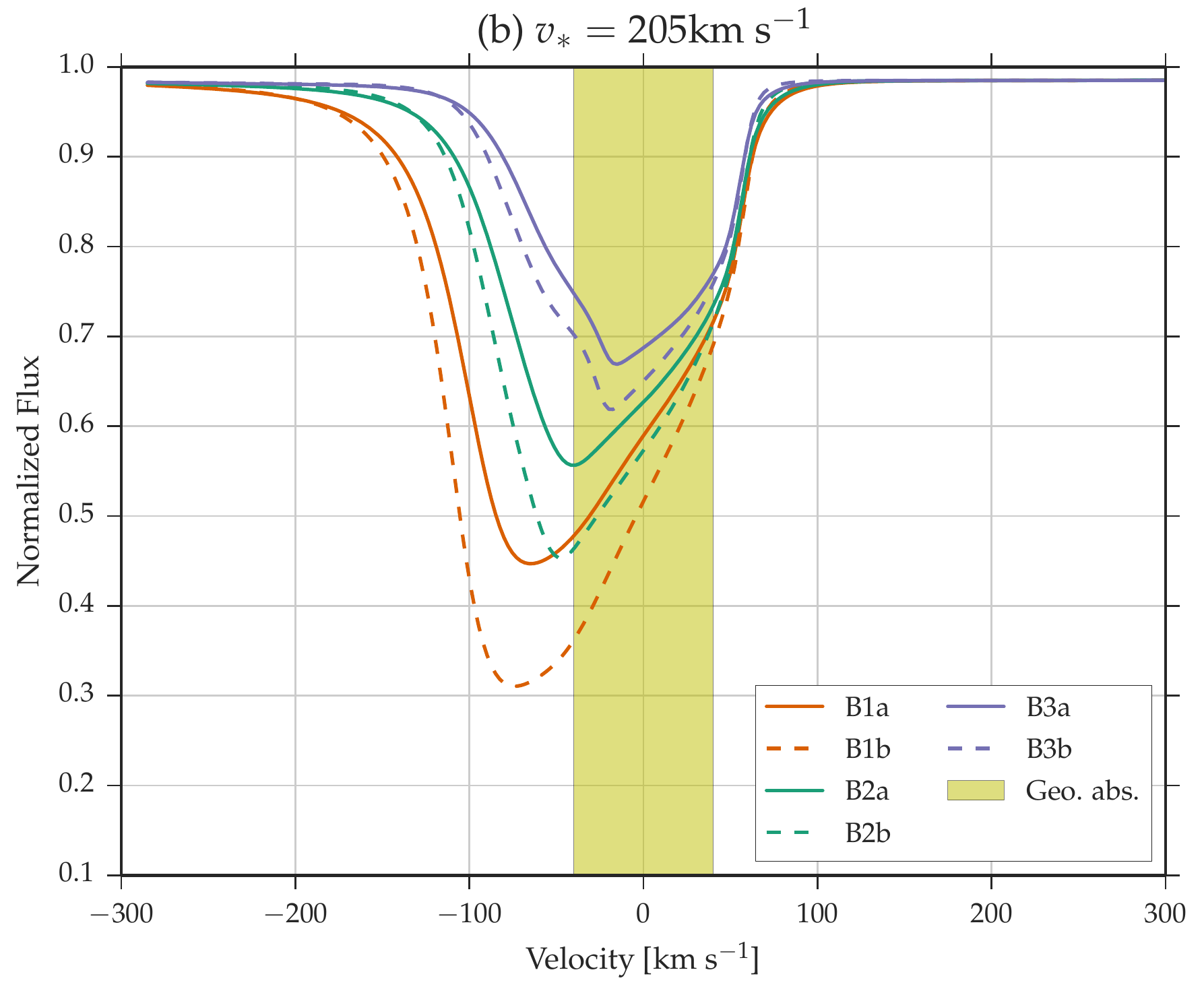}
  \end{subfigure}
  \begin{subfigure}{0.33\textwidth}
    \includegraphics[width=\textwidth]{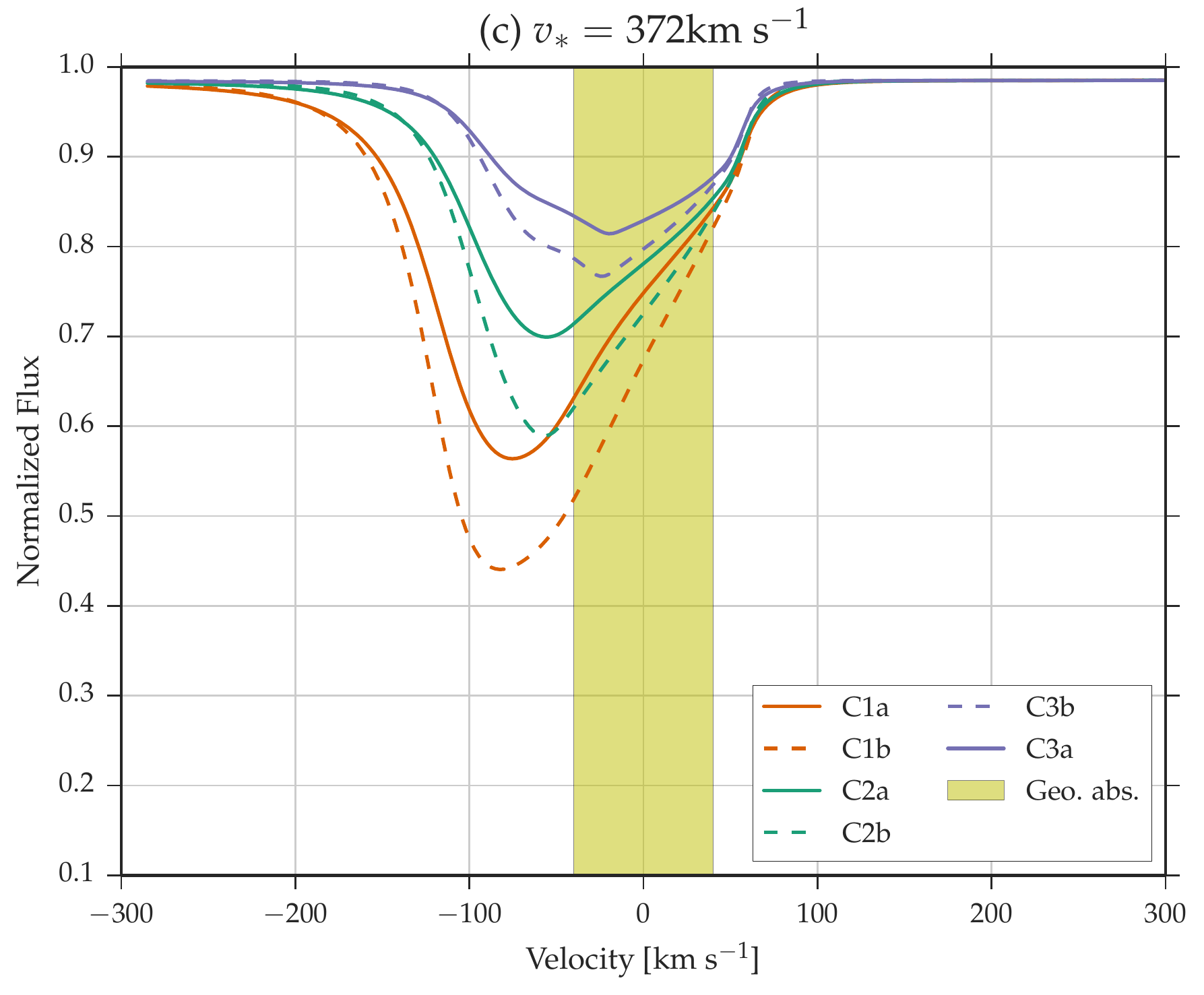}
  \end{subfigure}
  \caption{Normalized stellar emission at $t=2.9~\mathrm{days}$ as a
    function of LOS velocity. The models are grouped in the
    same manner as in Figure \ref{fig:totabs}
    Panel (a) includes the line profile for a
    simulations without radiation transfer (black line), calculated
    with the passive scalar and the a ionization temperature as in
    \citet{schneiter2007, villarreal2014}. The yellow
    stripe represents the part of the line that is contaminated with
    the geocoronal glow and was omitted form  the total absorption
    calculation.}\label{fig:velprof}
\end{figure*}

From the detection made with the Space Telescope Imaging
Spectrograph (STIS) on board of the Hubble Space Telescope (HST) in
2001 \citep[see][]{vidal2003} it is clear that the deepest absorption
signatures are found in the blue wing of the Ly-$\alpha$ line, in
the velocity range between $-130$ and
$-40~\mathrm{km~s^{-1}}$. Whereas the red side of the Doppler Shifted
absorption, defined within the velocity range between $32$ and
$100~\mathrm{km~s^{-1}}$ is less significant.
To study this features, we show in Figure \ref{fig:velprof}
the stellar emission as a function of velocity, for all the models, at
the time of maximum absorption during the second transit
($t=2.9~\mathrm{days}$).

From Figure \ref{fig:velprof} we notice that the absorption depends
strongly on the ionizing flux, especially on the blue side of the
spectrum, contrary to what it is observed on the red side, where all
models show similar behaviour.

 It is expected that higher ionizing fluxes result in a lower
 absorption, as photoionized material becomes transparent to
 Ly-$\alpha$. In addition, neutral material in the wake of the
 planetary wind accelerates due to its hydrodynamical interaction with
 the stellar wind, as well as due to radiation pressure (modeled here
 as a deficit in the gravitational pull of the star). The neutral wake
 therefore accelerates towards the observer (see Figure
 \ref{fig:cuts}), and interacts with the Ly-$\alpha$ emission from the
 star producing the prominent blue-shifted absorption,
   whose maximum value shifts towards more negatives velocities when
   the ionizing flux diminish.

\begin{figure*}
\centering
  \begin{subfigure}{0.33\textwidth}
    \includegraphics[width=\textwidth]{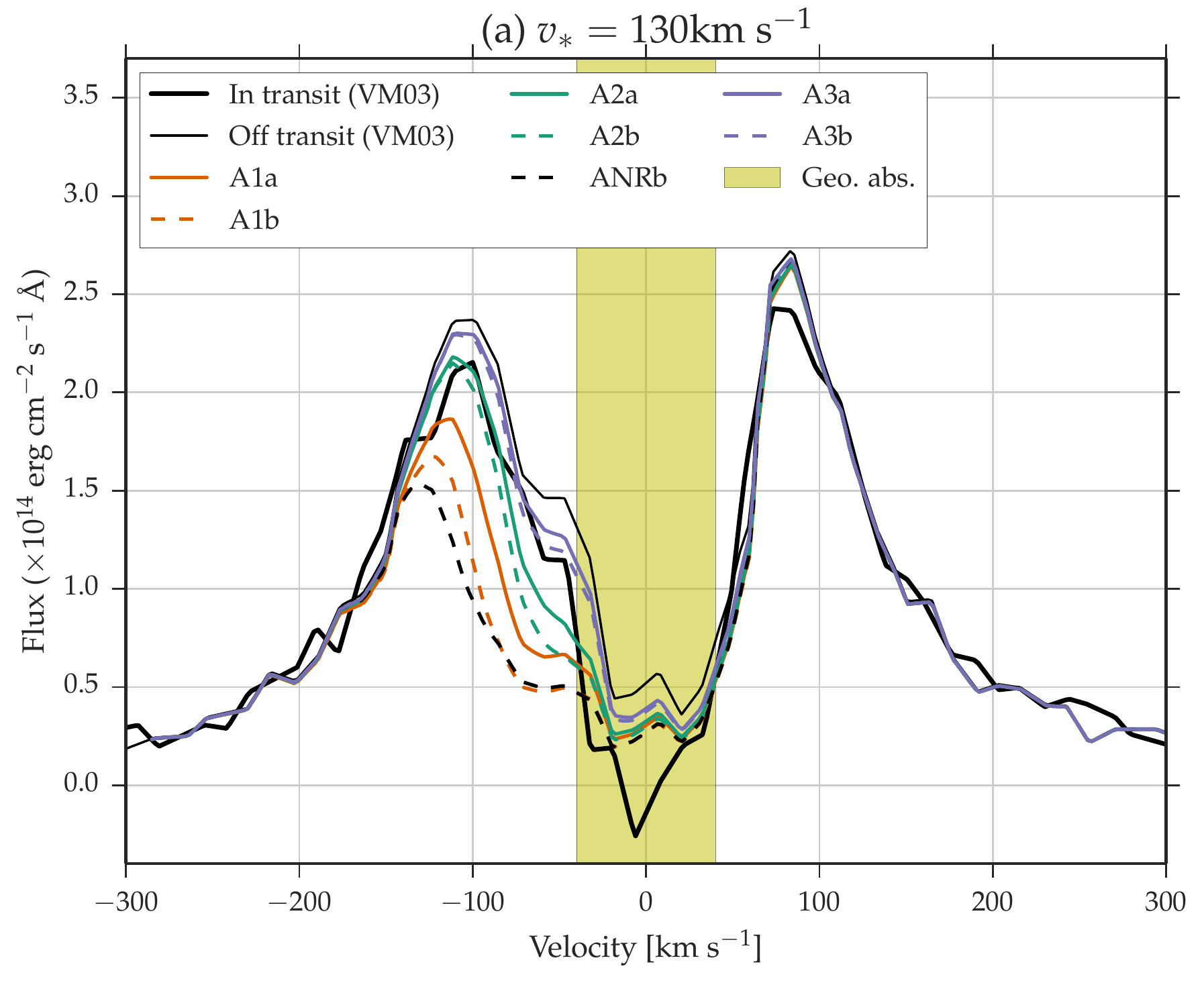}
  \end{subfigure}%
  \begin{subfigure}{0.33\textwidth}
    \includegraphics[width=\textwidth]{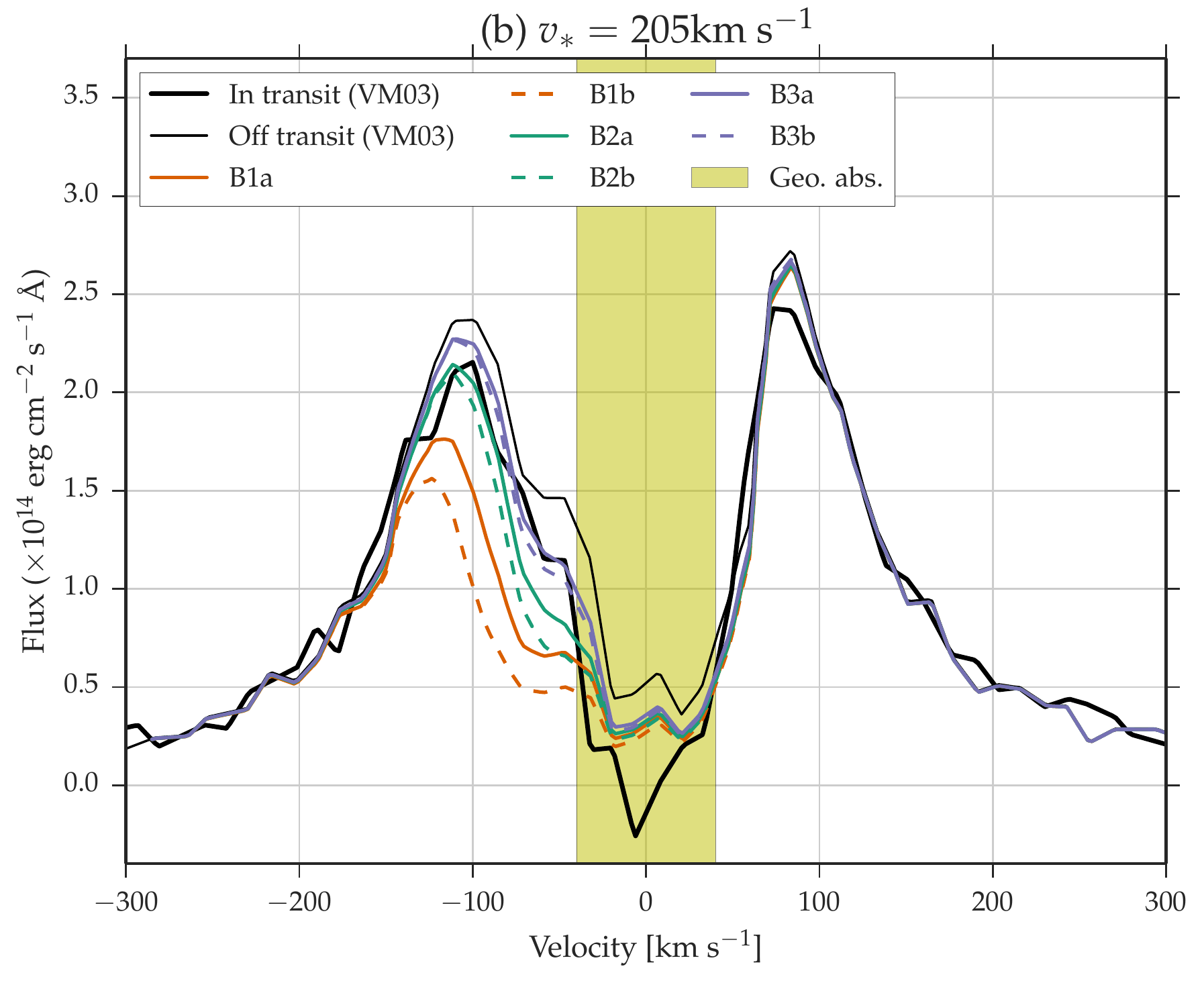}
  \end{subfigure}%
  \begin{subfigure}{0.33\textwidth}
    \includegraphics[width=\textwidth]{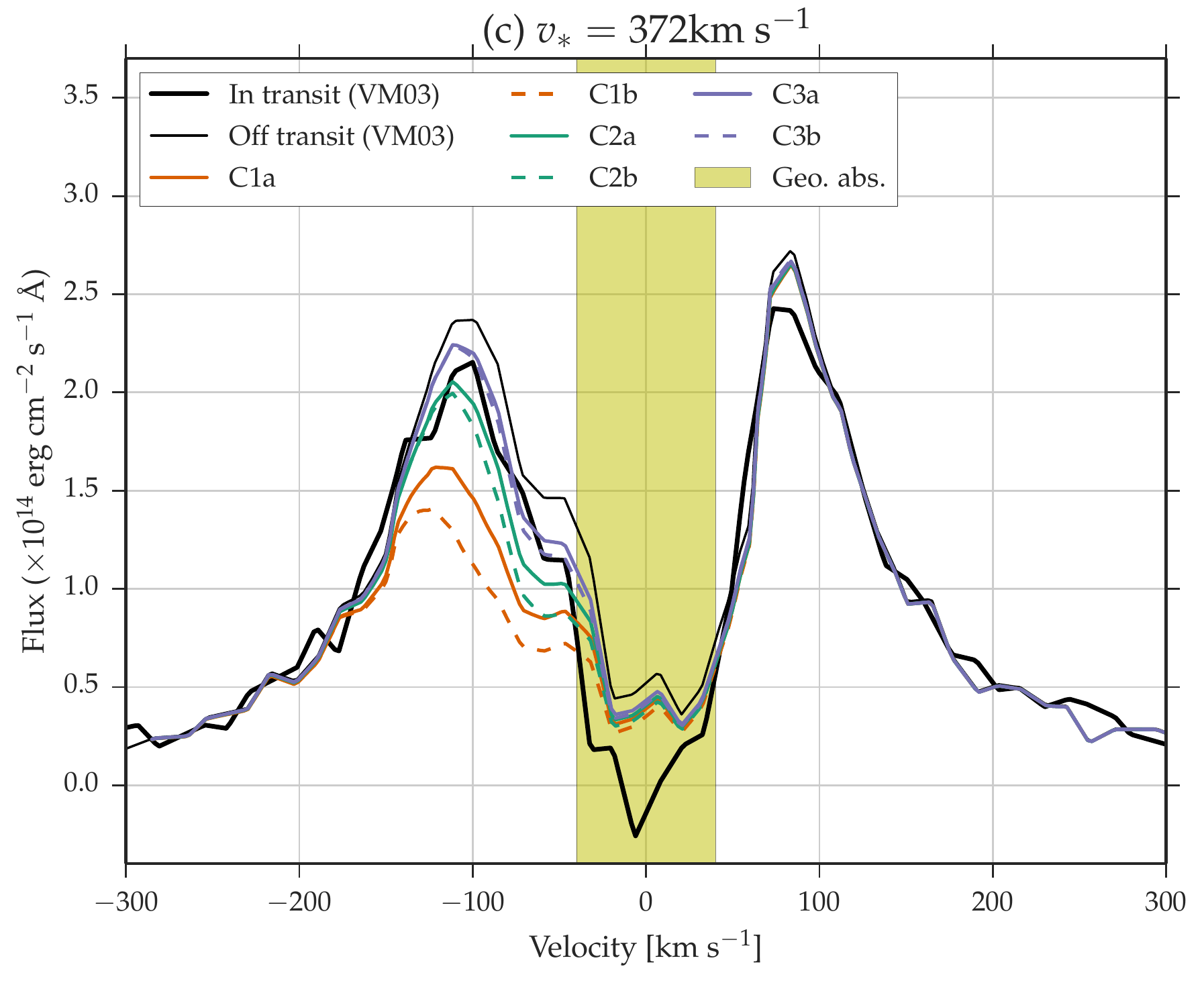}
  \end{subfigure}
  \caption{Ly-$\alpha$ observations \citep[reproduced
    from][]{vidal2003}, and the corresponding  synthetic absorption
    calculated with our models. The color and line coding is the same
    as in previous figures, and it is indicated in the label.}
\label{fig:obscomp}
\end{figure*}

To compare our results with the observations we took the original
off-transit data \citep[reproduced from][]{vidal2003}, and multiplied
them by the velocity dependent absorption factor in our models. The
resulting Ly-$\alpha$ profiles can be compared with the in-transit
observations in Figure \ref{fig:obscomp} \citep[also reproduced
  from][]{vidal2003}.

The observations show absorption in both the blue part of the spectrum
(in the velocity range from $-130$ to $-40~\mathrm{km~s^{-1}}$), as
well as in the red part of the spectrum (velocity range from $32$ to
$100~\mathrm{km~s^{-1}}$) with an attenuation of $\sim 10 \%$ and
$\sim 5\%$, respectively.
%

Most of the models reproduce the level of absorption in the blue wing
(except models with low stellar flux that overestimate the absorption
for all velocities), considering the large uncertainties in the
observations. The red wing absorption is generally underestimated, and
only a couple of models  show very little absorption around $90~\mathrm{km~s^{-1}}$.

In Figures \ref{fig:velprof} and \ref{fig:obscomp} the
  region shaded in yellow corresponds to the part of the spectrum
  contaminated by geocoronal emission in observations, thus it should
  not be considered when comparing our models with observations.

\section{Discussion}\label{discussion}

The models presented here are closer to a self-consistent study of the
stellar-planetary wind interaction than the two previous works
\citep{schneiter2007, villarreal2014}, where a proper photoionizating
photon flux was not included. 

Focusing on the the outer part of the planetary atmosphere ($r \gtrsim
3 R_\mathrm{p}$), and adjusting the hydrodynamic parameters according
the results in \citet{schneiter2007, villarreal2014}, we have done a
study of the influence of the high energy photon flux on the escaping
part of the atmosphere. The maximum absorption is shown in Figure
\ref{fig:totabs} (the portion shaded in green which corresponds to
the planet transit), where a strong almost linear dependence on the
high ionizing flux  is found. The importance of the planetary mass
loss rate becomes less obvious (almost negligible) within the accepted
range estimated in previous works
\citep{schneiter2007,villarreal2014}.  At the same time,
$\dot{M}_\mathrm{p}$ does affect the absorption in the tail (after the
planetary transit).  This result is not surprising, since most of the
neutral material is swept, by the stellar wind, to produce the tail
which, due to the orbital motion, and gravitational pull, it becomes curved
 getting exposed to the photoionizing flux incoming from the star (see
Figures \ref{fig:render} and \ref{fig:cuts}).

  The stellar EUV flux in our models ranges between
  $\sim [1 \times 10^2 - 4\times 10^3]$ erg $cm^{-2}$ $s^{-1}$. This
  implies that the planetary evaporation for HD 209458b will be driven by an energy
  limited mechanism \citep[see][]{murray-clay2009}. Under this paradigm we can estimate the
  planetary mass loss rate from the simple equation \citep{yelle2004,penz2008}
\begin{equation}
\dot{M}_p = \eta_\mathrm{rad}\frac{\pi R_\mathrm{p}^3 F_\mathrm{EUV}}{GM_\mathrm{p}}
\label{eqmp}
\end{equation}
where $ \eta_{rad}$ is the heating efficiency and $F_\mathrm{EUV}$ is the EUV
flux at the position of the planet. $M_\mathrm{p}$, $R_\mathrm{p}$ are the mass and the
planetary radius and $G$ is the gravitational constant. This formula
does not take into account the correction for the Roche lobe effects
that acording to \citet{penz2008, erkaev2007}, can change
the value of $\dot{M}_\mathrm{p}$ in a factor of $\sim 1.5$ for HD
209458b. \\ In our simulations, we fix the mass loss rate of the
planet and the stellar flux for each model. This implies, according to
the Eq. \ref{eqmp}, to assume a certain value for the heating efficiency at the
position where the planetary wind is launched. Hence, in our models,
the heating efficiency varies from 2$\%$ to 100 $\%$. At the present,
it is unclear what is the best value for $\eta$, but with our models we
will be covering almost all possibles values.  In the literature, several
authors tried to constrain this factor: \citet{tian2005} adopted a
$\eta=0.6$ for a pure atomic H atmosphere, \citet{ehrenreich2011}
obtained an $\eta > 1$ based on the values of mass loss rate from
\citet{vidal2003} and \citet{linsky2010} and the luminosity
value from \citet{kashyap2008}, \citet{lammer2009} suggest $\eta
\sim [10-25] \%$.\\
As a first aproximation, from Fig. \ref{fig:obscomp}, we can say that the lowest value of $F_0$ can't reproduce the observed absorption, 
hence this models are not expected to be representative of the planetary system parameters. So, the highest values of $\eta$ ($\sim [50-100] \%$) are not expected to define the planetary atmosphere of HD 209458b.\\
It would be important to have some observational constraint to the value of
 $\eta$, as mentioned \citet{ehrenreich2011}. This could shed some light into the atmospheric
 composition of exoplanets.

When comparing our models with the observations, in particular
focusing on the Doppler blue-shifted part of the curve (Figure
\ref{fig:obscomp}) we see that the $v_*=130~\mathrm{km~s^{-1}}$ peak
is better fitted with the $S_0$ photon rate, whereas a lower absolute
velocity part of the spectrum (between $-40$ to $-75~\mathrm{km~s^{-1}}$) 
is better fitted with a larger photon
rate ($5 \times S_0$) flux.  The same trend is observed for larger
wind velocity. Certainly with a more fine parameter coverage one could
reproduce the absorption in more detail, however this is beyond the
scope of this paper. 

In a recent work, \citet{tremblin2013} study the effect of charge
exchange in the mixing region of the two winds. With their choice of
parameters they are able to reproduce an asymmetric absorption
(also favouring the blue wing), in particular the $\sim 10 \%$
absorption at $-100~\mathrm{km~s^{-1}}$.
Their parameters are similar to our model A1b, but with the inclusion
of a treatment for charge exchange between the stellar and planetary
winds. 

In contrast to the studies of \citet{tremblin2013, holmstrom2008}, our
simulations did not inlude charge exchange but where still able to
produce absorption at high blue Doppler-shift velocities, but due to
the pure hydrodynamic interaction, along with the  global radiation
pressure\footnote{implemented as a reduction of the stellar
  gravitational potential as in
  \citet{vidal2003,schneiter2007,villarreal2014}.}. 
It is clear that both effects are expected to be present, and charge
exchange in models as ours would enhance the absorption in the blue.
An assessment of their relative (or collective) contribution is left
for a follow up study.

The inclusion of the radiation transfer calculation also shows that
the previous estimates of the planetary mass loss rates where a bit low. 
It is clear from Figure \ref{fig:totabs} that the same hydrodynamical
parameters result in a overestimation of the absorption respect to the
models that do include the photionization of the tail. Therefore, to
reach a similar absorption with the photonionization included requires
a larger mass loss rate. However the mass loss rates used in this work are
compatible to previous estimates.

\section{Conclusions}
\label{conclusions}

We used the 3D numerical hydrodynamic code {\sc Guacho}, to simulate
the  interaction of the wind escaping a hot-Jupiter (with parameters
resembling HD 209458b) and a photo-ionized wind coming from a
solar-type star (HD 209458).

The ionizing flux was found to have a great influence on the neutral 
hydrogen escaping from the planet, producing an almost linear effect 
on the absorption of the wake. 
With planetary mass loss rates within the range proposed in \citet{schneiter2007,
  villarreal2014} it is possible to find EUV fluxes that reproduce the
observed absorption measured in \citet{vidal2003} down to velocities
less than $-100~\mathrm{km~s^{-1}}$. 
Perfoming 3D simulations that include the orbital motion of the
planet is a natural way of obtaining a non-symmetric hydrogen envelop
that is subject to the incoming stellar photons and wind. 


Our models show that the hydrodynamic interaction
between the planetary and stellar wind is able to reproduce the
asymmetric absorption towards high blue-shifted velocities, as observed
in \citep{vidal2003}.
Such asymmetric absorption has been also attributed to charge-exchange
between the fast stellar wind and the slow planetary
wind \citep{tremblin2013}.

In a subsequent paper we plan to include the effects of
the stellar magnetic fields together with the charge-exchange
interaction. This will allow us to further constraint the photon
flux and its influence on the acceleration of the neutral
escaping material.


\section*{Acknowledgements}

MS acknowledges the financial support from the PICT project 2566/OC-AR and the
fellowship grant `BECA EXTERNA AGOSTO 2013' given by CONICET. 
AE, ACR and PFV acknowledge support from DGAPA-PAPIIT (UNAM) grants IN109715, IG100214, and CONACYT grant
167611. Authors also thank for financial support from CONACYT-CONICET joint grant CAR 190489.

\bibliographystyle{mnras}
\bibliography{ref}{}

\begin{thebibliography}{}
\makeatletter
\relax
\def\mn@urlcharsother{\let\do\@makeother \do\$\do\&\do\#\do\^\do\_\do\%\do\~}
\def\mn@doi{\begingroup\mn@urlcharsother \@ifnextchar [ {\mn@doi@}
  {\mn@doi@[]}}
\def\mn@doi@[#1]#2{\def\@tempa{#1}\ifx\@tempa\@empty \href
  {http://dx.doi.org/#2} {doi:#2}\else \href {http://dx.doi.org/#2} {#1}\fi
  \endgroup}
\def\mn@eprint#1#2{\mn@eprint@#1:#2::\@nil}
\def\mn@eprint@arXiv#1{\href {http://arxiv.org/abs/#1} {{\tt arXiv:#1}}}
\def\mn@eprint@dblp#1{\href {http://dblp.uni-trier.de/rec/bibtex/#1.xml}
  {dblp:#1}}
\def\mn@eprint@#1:#2:#3:#4\@nil{\def\@tempa {#1}\def\@tempb {#2}\def\@tempc
  {#3}\ifx \@tempc \@empty \let \@tempc \@tempb \let \@tempb \@tempa \fi \ifx
  \@tempb \@empty \def\@tempb {arXiv}\fi \@ifundefined
  {mn@eprint@\@tempb}{\@tempb:\@tempc}{\expandafter \expandafter \csname
  mn@eprint@\@tempb\endcsname \expandafter{\@tempc}}}

\bibitem[\protect\citeauthoryear{{Adams}}{{Adams}}{2011}]{adams2011}
{Adams} F.~C.,  2011, \mn@doi [\apj] {10.1088/0004-637X/730/1/27}, \href
  {http://adsabs.harvard.edu/abs/2011ApJ...730...27A} {730, 27}

\bibitem[\protect\citeauthoryear{{Baraffe}, {Chabrier}, {Barman}, {Selsis},
  {Allard}  \& {Hauschildt}}{{Baraffe} et~al.}{2005}]{baraffe2005}
{Baraffe} I.,  {Chabrier} G.,  {Barman} T.~S.,  {Selsis} F.,  {Allard} F.,
  {Hauschildt} P.~H.,  2005, \mn@doi [aap] {10.1051/0004-6361:200500123}, \href
  {http://adsabs.harvard.edu/abs/2005A\%26A...436L..47B} {436, L47}

\bibitem[\protect\citeauthoryear{{Batygin} \& {Stanley}}{{Batygin} \&
  {Stanley}}{2014}]{batygin2014}
{Batygin} K.,  {Stanley} S.,  2014, \mn@doi [ApJ] {10.1088/0004-637X/794/1/10},
  \href {http://adsabs.harvard.edu/abs/2014ApJ...794...10B} {794, 10}

\bibitem[\protect\citeauthoryear{{Ben-Jaffel}}{{Ben-Jaffel}}{2007}]{benjaffel2007}
{Ben-Jaffel} L.,  2007, \mn@doi [ApJL] {10.1086/524706}, \href
  {http://adsabs.harvard.edu/abs/2007ApJ...671L..61B} {671, L61}

\bibitem[\protect\citeauthoryear{{Ben-Jaffel}}{{Ben-Jaffel}}{2008}]{benjaffel2008}
{Ben-Jaffel} L.,  2008, \mn@doi [ApJ] {10.1086/592101}, \href
  {http://adsabs.harvard.edu/abs/2008ApJ...688.1352B} {688, 1352}

\bibitem[\protect\citeauthoryear{{Bourrier} \& {Lecavelier des
  Etangs}}{{Bourrier} \& {Lecavelier des Etangs}}{2013}]{bourrier2013}
{Bourrier} V.,  {Lecavelier des Etangs} A.,  2013, \mn@doi [A\&A]
  {10.1051/0004-6361/201321551}, \href
  {http://adsabs.harvard.edu/abs/2013A%26A...557A.124B} {557, A124}

\bibitem[\protect\citeauthoryear{{Bourrier}, {Ehrenreich}  \& {Lecavelier des
  Etangs}}{{Bourrier} et~al.}{2015}]{bourrier2015}
{Bourrier} V.,  {Ehrenreich} D.,   {Lecavelier des Etangs} A.,  2015, \mn@doi
  [A\&A] {10.1051/0004-6361/201526894}, \href
  {http://adsabs.harvard.edu/abs/2015A%26A...582A..65B} {582, A65}

\bibitem[\protect\citeauthoryear{{Charbonneau}, {Brown}, {Latham}  \&
  {Mayor}}{{Charbonneau} et~al.}{2000}]{charbonneau2000}
{Charbonneau} D.,  {Brown} T.~M.,  {Latham} D.~W.,   {Mayor} M.,  2000, \mn@doi
  [ApjL] {10.1086/312457}, \href
  {http://adsabs.harvard.edu/abs/2000ApJ...529L..45C} {529, L45}

\bibitem[\protect\citeauthoryear{{Cohen}, {Kashyap}, {Drake}, {Sokolov},
  {Garraffo}  \& {Gombosi}}{{Cohen} et~al.}{2011a}]{cohen2011a}
{Cohen} O.,  {Kashyap} V.~L.,  {Drake} J.~J.,  {Sokolov} I.~V.,  {Garraffo} C.,
    {Gombosi} T.~I.,  2011a, \mn@doi [\apj] {10.1088/0004-637X/733/1/67}, \href
  {http://adsabs.harvard.edu/abs/2011ApJ...733...67C} {733, 67}

\bibitem[\protect\citeauthoryear{{Cohen}, {Kashyap}, {Drake}, {Sokolov}  \&
  {Gombosi}}{{Cohen} et~al.}{2011b}]{cohen2011b}
{Cohen} O.,  {Kashyap} V.~L.,  {Drake} J.~J.,  {Sokolov} I.~V.,   {Gombosi}
  T.~I.,  2011b, \mn@doi [\apj] {10.1088/0004-637X/738/2/166}, \href
  {http://adsabs.harvard.edu/abs/2011ApJ...738..166C} {738, 166}

\bibitem[\protect\citeauthoryear{{Ehrenreich} \& {D{\'e}sert}}{{Ehrenreich} \&
  {D{\'e}sert}}{2011}]{ehrenreich2011}
{Ehrenreich} D.,  {D{\'e}sert} J.-M.,  2011, \mn@doi [A\&A]
  {10.1051/0004-6361/201016356}, \href
  {http://adsabs.harvard.edu/abs/2011A%26A...529A.136E} {529, A136}

\bibitem[\protect\citeauthoryear{{Ekenb{\"a}ck}, {Holmstr{\"o}m}, {Wurz},
  {Grie{\ss}meier}, {Lammer}, {Selsis}  \& {Penz}}{{Ekenb{\"a}ck}
  et~al.}{2010}]{ekenback2010}
{Ekenb{\"a}ck} A.,  {Holmstr{\"o}m} M.,  {Wurz} P.,  {Grie{\ss}meier} J.-M.,
  {Lammer} H.,  {Selsis} F.,   {Penz} T.,  2010, \mn@doi [ApJ]
  {10.1088/0004-637X/709/2/670}, \href
  {http://adsabs.harvard.edu/abs/2010ApJ...709..670E} {709, 670}

\bibitem[\protect\citeauthoryear{{Erkaev}, {Kulikov}, {Lammer}, {Selsis},
  {Langmayr}, {Jaritz}  \& {Biernat}}{{Erkaev} et~al.}{2007}]{erkaev2007}
{Erkaev} N.~V.,  {Kulikov} Y.~N.,  {Lammer} H.,  {Selsis} F.,  {Langmayr} D.,
  {Jaritz} G.~F.,   {Biernat} H.~K.,  2007, \mn@doi [A\& A]
  {10.1051/0004-6361:20066929}, \href
  {http://adsabs.harvard.edu/abs/2007A%26A...472..329E} {472, 329}

\bibitem[\protect\citeauthoryear{{Esquivel} \& {Raga}}{{Esquivel} \&
  {Raga}}{2013}]{2013ApJ...779..111E}
{Esquivel} A.,  {Raga} A.~C.,  2013, \mn@doi [ApJ]
  {10.1088/0004-637X/779/2/111}, \href
  {http://adsabs.harvard.edu/abs/2013ApJ...779..111E} {779, 111}

\bibitem[\protect\citeauthoryear{{Esquivel}, {Raga}, {Cant{\'o}}  \&
  {Rodr{\'{\i}}guez-Gonz{\'a}lez}}{{Esquivel} et~al.}{2009}]{esquivel2009}
{Esquivel} A.,  {Raga} A.~C.,  {Cant{\'o}} J.,
  {Rodr{\'{\i}}guez-Gonz{\'a}lez} A.,  2009, \mn@doi [A\&A]
  {10.1051/0004-6361/200912825}, \href
  {http://adsabs.harvard.edu/abs/2009A%26A...507..855E} {507, 855}

\bibitem[\protect\citeauthoryear{{Garc{\'{\i}}a Mu{\~n}oz}}{{Garc{\'{\i}}a
  Mu{\~n}oz}}{2007}]{garcia2007}
{Garc{\'{\i}}a Mu{\~n}oz} A.,  2007, \mn@doi [planss]
  {10.1016/j.pss.2007.03.007}, \href
  {http://adsabs.harvard.edu/abs/2007P%26SS...55.1426G} {55, 1426}

\bibitem[\protect\citeauthoryear{{Hartigan} \& {Raymond}}{{Hartigan} \&
  {Raymond}}{1993}]{1993ApJ...409..705H}
{Hartigan} P.,  {Raymond} J.,  1993, \mn@doi [ApJ] {10.1086/172700}, \href
  {http://adsabs.harvard.edu/abs/1993ApJ...409..705H} {409, 705}

\bibitem[\protect\citeauthoryear{{Henry}, {Marcy}, {Butler}  \& {Vogt}}{{Henry}
  et~al.}{2000}]{henry2000}
{Henry} G.~W.,  {Marcy} G.~W.,  {Butler} R.~P.,   {Vogt} S.~S.,  2000, \mn@doi
  [ApJL] {10.1086/312458}, \href
  {http://adsabs.harvard.edu/abs/2000ApJ...529L..41H} {529, L41}

\bibitem[\protect\citeauthoryear{{Holmstr{\"o}m}, {Ekenb{\"a}ck}, {Selsis},
  {Penz}, {Lammer}  \& {Wurz}}{{Holmstr{\"o}m} et~al.}{2008}]{holmstrom2008}
{Holmstr{\"o}m} M.,  {Ekenb{\"a}ck} A.,  {Selsis} F.,  {Penz} T.,  {Lammer} H.,
    {Wurz} P.,  2008, \mn@doi [NATURE] {10.1038/nature06600}, \href
  {http://adsabs.harvard.edu/abs/2008Natur.451..970H} {451, 970}

\bibitem[\protect\citeauthoryear{{Kashyap}, {Drake}  \& {Saar}}{{Kashyap}
  et~al.}{2008}]{kashyap2008}
{Kashyap} V.~L.,  {Drake} J.~J.,   {Saar} S.~H.,  2008, \mn@doi [ApJ]
  {10.1086/591922}, \href {http://adsabs.harvard.edu/abs/2008ApJ...687.1339K}
  {687, 1339}

\bibitem[\protect\citeauthoryear{{Koskinen}, {Harris}, {Yelle}  \&
  {Lavvas}}{{Koskinen} et~al.}{2013}]{koskinen2013}
{Koskinen} T.~T.,  {Harris} M.~J.,  {Yelle} R.~V.,   {Lavvas} P.,  2013,
  \mn@doi [ICARUS] {10.1016/j.icarus.2012.09.027}, \href
  {http://adsabs.harvard.edu/abs/2013Icar..226.1678K} {226, 1678}

\bibitem[\protect\citeauthoryear{{Lammer}, {Selsis}, {Ribas}, {Guinan}, {Bauer}
   \& {Weiss}}{{Lammer} et~al.}{2003}]{lammer2003}
{Lammer} H.,  {Selsis} F.,  {Ribas} I.,  {Guinan} E.~F.,  {Bauer} S.~J.,
  {Weiss} W.~W.,  2003, \mn@doi [ApJL] {10.1086/380815}, \href
  {http://adsabs.harvard.edu/abs/2003ApJ...598L.121L} {598, L121}

\bibitem[\protect\citeauthoryear{{Lammer} et~al.,}{{Lammer}
  et~al.}{2009}]{lammer2009}
{Lammer} H.,  et~al., 2009, \mn@doi [A\&A] {10.1051/0004-6361/200911922}, \href
  {http://adsabs.harvard.edu/abs/2009A%26A...506..399L} {506, 399}

\bibitem[\protect\citeauthoryear{{Lecavelier des Etangs}, {Vidal-Madjar},
  {McConnell}  \& {H{\'e}brard}}{{Lecavelier des Etangs}
  et~al.}{2004}]{lecavelier2004}
{Lecavelier des Etangs} A.,  {Vidal-Madjar} A.,  {McConnell} J.~C.,
  {H{\'e}brard} G.,  2004, \mn@doi [A\&A] {10.1051/0004-6361:20040106}, \href
  {http://adsabs.harvard.edu/abs/2004A%26A...418L...1L} {418, L1}

\bibitem[\protect\citeauthoryear{{Linsky}, {Yang}, {France}, {Froning},
  {Green}, {Stocke}  \& {Osterman}}{{Linsky} et~al.}{2010}]{linsky2010}
{Linsky} J.~L.,  {Yang} H.,  {France} K.,  {Froning} C.~S.,  {Green} J.~C.,
  {Stocke} J.~T.,   {Osterman} S.~N.,  2010, \mn@doi [\apj]
  {10.1088/0004-637X/717/2/1291}, \href
  {http://adsabs.harvard.edu/abs/2010ApJ...717.1291L} {717, 1291}

\bibitem[\protect\citeauthoryear{{Matsakos}, {Uribe}  \&
  {K{\"o}nigl}}{{Matsakos} et~al.}{2015}]{matsakos2015}
{Matsakos} T.,  {Uribe} A.,   {K{\"o}nigl} A.,  2015, \mn@doi [A\&A]
  {10.1051/0004-6361/201425593}, \href
  {http://adsabs.harvard.edu/abs/2015A%26A...578A...6M} {578, A6}

\bibitem[\protect\citeauthoryear{{Murray-Clay}, {Chiang}  \&
  {Murray}}{{Murray-Clay} et~al.}{2009}]{murray-clay2009}
{Murray-Clay} R.~A.,  {Chiang} E.~I.,   {Murray} N.,  2009, \mn@doi [ApJ.]
  {10.1088/0004-637X/693/1/23}, \href
  {http://adsabs.harvard.edu/abs/2009ApJ...693...23M} {693, 23}

\bibitem[\protect\citeauthoryear{{Owen} \& {Adams}}{{Owen} \&
  {Adams}}{2014}]{owen2014}
{Owen} J.~E.,  {Adams} F.~C.,  2014, \mn@doi [\mnras] {10.1093/mnras/stu1684},
  \href {http://adsabs.harvard.edu/abs/2014MNRAS.444.3761O} {444, 3761}

\bibitem[\protect\citeauthoryear{{Owen} \& {Jackson}}{{Owen} \&
  {Jackson}}{2012}]{owen2012}
{Owen} J.~E.,  {Jackson} A.~P.,  2012, \mn@doi [MNRAS]
  {10.1111/j.1365-2966.2012.21481.x}, \href
  {http://adsabs.harvard.edu/abs/2012MNRAS.425.2931O} {425, 2931}

\bibitem[\protect\citeauthoryear{{Penz} et~al.,}{{Penz}
  et~al.}{2008}]{penz2008}
{Penz} T.,  et~al., 2008, \mn@doi [\planss] {10.1016/j.pss.2008.04.005}, \href
  {http://adsabs.harvard.edu/abs/2008P%26SS...56.1260P} {56, 1260}

\bibitem[\protect\citeauthoryear{{Sanz-Forcada}, {Micela}, {Ribas}, {Pollock},
  {Eiroa}, {Velasco}, {Solano}  \& {Garc{\'{\i}}a-{\'A}lvarez}}{{Sanz-Forcada}
  et~al.}{2011}]{sanz-forcada2011}
{Sanz-Forcada} J.,  {Micela} G.,  {Ribas} I.,  {Pollock} A.~M.~T.,  {Eiroa} C.,
   {Velasco} A.,  {Solano} E.,   {Garc{\'{\i}}a-{\'A}lvarez} D.,  2011, \mn@doi
  [A\&A] {10.1051/0004-6361/201116594}, \href
  {http://adsabs.harvard.edu/abs/2011A%26A...532A...6S} {532, A6}

\bibitem[\protect\citeauthoryear{{Schneiter}, {Vel{\'a}zquez}, {Esquivel},
  {Raga}  \& {Blanco-Cano}}{{Schneiter} et~al.}{2007}]{schneiter2007}
{Schneiter} E.~M.,  {Vel{\'a}zquez} P.~F.,  {Esquivel} A.,  {Raga} A.~C.,
  {Blanco-Cano} X.,  2007, \mn@doi [ApJL] {10.1086/524945}, \href
  {http://adsabs.harvard.edu/abs/2007ApJ...671L..57S} {671, L57}

\bibitem[\protect\citeauthoryear{{Showman}, {Cooper}, {Fortney}  \&
  {Marley}}{{Showman} et~al.}{2008}]{showman2008}
{Showman} A.~P.,  {Cooper} C.~S.,  {Fortney} J.~J.,   {Marley} M.~S.,  2008,
  \mn@doi [ApJ.] {10.1086/589325}, \href
  {http://adsabs.harvard.edu/abs/2008ApJ...682..559S} {682, 559}

\bibitem[\protect\citeauthoryear{{Tian}, {Toon}, {Pavlov}  \& {De
  Sterck}}{{Tian} et~al.}{2005}]{tian2005}
{Tian} F.,  {Toon} O.~B.,  {Pavlov} A.~A.,   {De Sterck} H.,  2005, \mn@doi
  [ApJ] {10.1086/427204}, \href
  {http://adsabs.harvard.edu/abs/2005ApJ...621.1049T} {621, 1049}

\bibitem[\protect\citeauthoryear{{Toro}}{{Toro}}{1999}]{torobook}
{Toro} E.~F.,  1999, {Riemann Solvers and Numerical Methods for Fluid
  Dynamics}.
Springer

\bibitem[\protect\citeauthoryear{{Trammell}, {Li}  \& {Arras}}{{Trammell}
  et~al.}{2014}]{trammell2014}
{Trammell} G.~B.,  {Li} Z.-Y.,   {Arras} P.,  2014, \mn@doi [\apj]
  {10.1088/0004-637X/788/2/161}, \href
  {http://adsabs.harvard.edu/abs/2014ApJ...788..161T} {788, 161}

\bibitem[\protect\citeauthoryear{{Tremblin} \& {Chiang}}{{Tremblin} \&
  {Chiang}}{2013}]{tremblin2013}
{Tremblin} P.,  {Chiang} E.,  2013, \mn@doi [MNRAS] {10.1093/mnras/sts212},
  \href {http://adsabs.harvard.edu/abs/2013MNRAS.428.2565T} {428, 2565}

\bibitem[\protect\citeauthoryear{{Tripathi}, {Kratter}, {Murray-Clay}  \&
  {Krumholz}}{{Tripathi} et~al.}{2015}]{tripathi2015}
{Tripathi} A.,  {Kratter} K.~M.,  {Murray-Clay} R.~A.,   {Krumholz} M.~R.,
  2015, \mn@doi [\apj] {10.1088/0004-637X/808/2/173}, \href
  {http://adsabs.harvard.edu/abs/2015ApJ...808..173T} {808, 173}

\bibitem[\protect\citeauthoryear{{Vidal-Madjar}, {Lecavelier des Etangs},
  {D{\'e}sert}, {Ballester}, {Ferlet}, {H{\'e}brard}  \&
  {Mayor}}{{Vidal-Madjar} et~al.}{2003}]{vidal2003}
{Vidal-Madjar} A.,  {Lecavelier des Etangs} A.,  {D{\'e}sert} J.-M.,
  {Ballester} G.~E.,  {Ferlet} R.,  {H{\'e}brard} G.,   {Mayor} M.,  2003,
  \mn@doi [Nature] {10.1038/nature01448}, \href
  {http://adsabs.harvard.edu/abs/2003Natur.422..143V} {422, 143}

\bibitem[\protect\citeauthoryear{{Vidal-Madjar}, {Lecavelier des Etangs},
  {D{\'e}sert}, {Ballester}, {Ferlet}, {H{\'e}brard}  \&
  {Mayor}}{{Vidal-Madjar} et~al.}{2008}]{vidal2008}
{Vidal-Madjar} A.,  {Lecavelier des Etangs} A.,  {D{\'e}sert} J.-M.,
  {Ballester} G.~E.,  {Ferlet} R.,  {H{\'e}brard} G.,   {Mayor} M.,  2008,
  \mn@doi [ApJL] {10.1086/587036}, \href
  {http://adsabs.harvard.edu/abs/2008ApJ...676L..57V} {676, L57}

\bibitem[\protect\citeauthoryear{{Villarreal D'Angelo}, {Schneiter}, {Costa},
  {Vel{\'a}zquez}, {Raga}  \& {Esquivel}}{{Villarreal D'Angelo}
  et~al.}{2014}]{villarreal2014}
{Villarreal D'Angelo} C.,  {Schneiter} M.,  {Costa} A.,  {Vel{\'a}zquez} P.,
  {Raga} A.,   {Esquivel} A.,  2014, \mn@doi [MNRAS] {10.1093/mnras/stt2303},
  \href {http://adsabs.harvard.edu/abs/2014MNRAS.438.1654V} {438, 1654}

\bibitem[\protect\citeauthoryear{{Yelle}}{{Yelle}}{2004}]{yelle2004}
{Yelle} R.~V.,  2004, \mn@doi [ICARUS] {10.1016/j.icarus.2004.02.008}, \href
  {http://adsabs.harvard.edu/abs/2004Icar..170..167Y} {170, 167}

\bibitem[\protect\citeauthoryear{{Yelle}}{{Yelle}}{2006}]{yelle2006}
{Yelle} R.~V.,  2006, \mn@doi [ICARUS] {10.1016/j.icarus.2006.05.001}, \href
  {http://adsabs.harvard.edu/abs/2006Icar..183..508Y} {183, 508}

\makeatother
\end{thebibliography}



\bsp
\label{lastpage}

\end{document}